\documentclass[%
 superscriptaddress,
 twocolumn,
 amsmath,amssymb,
 aps,
 prd,
floatfix,
]{revtex4-1}

\usepackage{graphicx}% Include figure files
\usepackage{dcolumn}% Align table columns on decimal point
\usepackage{bm}% bold math
\usepackage{amsmath}
\usepackage{mathptmx}
\usepackage{chemformula}
\begin{document}

\title{Determining the bubble nucleation efficiency of low-energy nuclear recoils in superheated C$_3$F$_8$ dark matter detectors}

\author{B.~Ali}
\affiliation{Institute of Experimental and Applied Physics, Czech Technical University in Prague, Prague, Cz-12800, Czech Republic}

\author{I.~J.~Arnquist}
\affiliation{Pacific Northwest National Laboratory, Richland, Washington 99354, USA}

\author{D.~Baxter}
\affiliation{Fermi National Accelerator Laboratory, Batavia, Illinois 60510, USA}

\author{E.~Behnke}
\affiliation{Department of Physics, Indiana University South Bend, South Bend, Indiana 46634, USA}

\author{M.~Bressler}
\affiliation{Department of Physics, Drexel University, Philadelphia, Pennsylvania 19104, USA}

\author{B.~Broerman}
\affiliation{Department of Physics, Queen's University, Kingston, K7L 3N6, Canada}

\author{K.~Clark}
\affiliation{Department of Physics, Queen's University, Kingston, K7L 3N6, Canada}

\author{J.~I.~Collar}
\affiliation{Enrico Fermi Institute, KICP and Department of Physics,
University of Chicago, Chicago, Illinois 60637, USA}

\author{P.~S.~Cooper}
\affiliation{Fermi National Accelerator Laboratory, Batavia, Illinois 60510, USA}

\author{C.~Cripe}
\affiliation{Department of Physics, Indiana University South Bend, South Bend, Indiana 46634, USA}

\author{M.~Crisler}
\affiliation{Fermi National Accelerator Laboratory, Batavia, Illinois 60510, USA}
\affiliation{Pacific Northwest National Laboratory, Richland, Washington 99354, USA}

\author{C.~E.~Dahl}
\affiliation{Department of Physics and Astronomy, Northwestern University, Evanston, Illinois 60208, USA}
\affiliation{Fermi National Accelerator Laboratory, Batavia, Illinois 60510, USA}

\author{M.~Das}
\affiliation{High Energy Nuclear \& Particle Physics Division, Saha Institute of Nuclear Physics, 1/AF, Bidhannagar Kolkata 700 064, India}

\author{D.~Durnford}
\email{ddurnfor@ualberta.ca}
\affiliation{Department of Physics, University of Alberta, Edmonton, T6G 2E1, Canada}

\author{S.~Fallows}
\affiliation{Department of Physics, University of Alberta, Edmonton, T6G 2E1, Canada}

\author{J.~Farine}
\affiliation{School of Natural Sciences, Laurentian University, Sudbury, P3E 2C6, Canada}
\affiliation{SNOLAB, Lively, Ontario, P3Y 1N2, Canada}
\affiliation{Department of Physics, Carleton University, Ottawa, Ontario, K1S 5B6, Canada}

\author{R.~Filgas}
\affiliation{Institute of Experimental and Applied Physics, Czech Technical University in Prague, Prague, Cz-12800, Czech Republic}

\author{A.~Garc\'{\i}a-Viltres}
\affiliation{Instituto de F\'isica, Universidad Nacional Aut\'onoma de M\'exico, A.P. 20-364, Ciudad de M\'exico 01000, M\'exico}

\author{F.~Girard}
\affiliation{School of Natural Sciences, Laurentian University, Sudbury, P3E 2C6, Canada}
\affiliation{D\'epartement de Physique, Universit\'e de Montr\'eal, Montr\'eal, H3C 3J7, Canada}

\author{G.~Giroux}
\affiliation{Department of Physics, Queen's University, Kingston, K7L 3N6, Canada}

\author{O.~Harris}
\affiliation{Northeastern Illinois University, Chicago, Illinois 60625, USA}

\author{E.~W.~Hoppe}
\affiliation{Pacific Northwest National Laboratory, Richland, Washington 99354, USA}

\author{C.~M.~Jackson}
\affiliation{Pacific Northwest National Laboratory, Richland, Washington 99354, USA}

\author{M.~Jin} 
\altaffiliation[now at ]{Centre for Artificial Intelligence and Robotics Hong Kong Institute of Science \& Innovation, Chinese Academy of Sciences 3/F, 17W, Science Park West Avenue, Hong Kong Science Park, Pak Shek Kok, New Territories, Hong Kong, China}
\affiliation{Department of Physics and Astronomy, Northwestern University, Evanston, Illinois 60208, USA}

\author{C.~B.~Krauss}
\affiliation{Department of Physics, University of Alberta, Edmonton, T6G 2E1, Canada}

\author{V.~Kumar}
\affiliation{High Energy Nuclear \& Particle Physics Division, Saha Institute of Nuclear Physics, Kolkata, India}

\author{M.~Lafreniere}
\affiliation{D\'epartement de Physique, Universit\'e de Montr\'eal, Montr\'eal, H3C 3J7, Canada}

\author{M.~Laurin}
\affiliation{D\'epartement de Physique, Universit\'e de Montr\'eal, Montr\'eal, H3C 3J7, Canada}

\author{I.~Lawson}
\affiliation{School of Natural Sciences, Laurentian University, Sudbury, P3E 2C6, Canada}
\affiliation{SNOLAB, Lively, Ontario, P3Y 1N2, Canada}

\author{A.~Leblanc}
\affiliation{School of Natural Sciences, Laurentian University, Sudbury, P3E 2C6, Canada}

\author{H.~Leng}
\affiliation{Materials Research Institute, Penn State, University Park, Pennsylvania 16802, USA}

\author{I.~Levine}
\affiliation{Department of Physics, Indiana University South Bend, South Bend, Indiana 46634, USA}

\author{C.~Licciardi}
\affiliation{School of Natural Sciences, Laurentian University, Sudbury, P3E 2C6, Canada}
\affiliation{SNOLAB, Lively, Ontario, P3Y 1N2, Canada}
\affiliation{Department of Physics, Carleton University, Ottawa, Ontario, K1S 5B6, Canada}

\author{S.~Linden}
\affiliation{SNOLAB, Lively, Ontario, P3Y 1N2, Canada}

\author{P.~Mitra}
\affiliation{Department of Physics, University of Alberta, Edmonton, T6G 2E1, Canada}

\author{V.~Monette}
\affiliation{D\'epartement de Physique, Universit\'e de Montr\'eal, Montr\'eal, H3C 3J7, Canada}

\author{C.~Moore}
\affiliation{Department of Physics, Queen's University, Kingston, K7L 3N6, Canada}

\author{R.~Neilson}
\affiliation{Department of Physics, Drexel University, Philadelphia, Pennsylvania 19104, USA}

\author{A.~J.~Noble}
\affiliation{Department of Physics, Queen's University, Kingston, K7L 3N6, Canada}

\author{H.~Nozard}
\affiliation{D\'epartement de Physique, Universit\'e de Montr\'eal, Montr\'eal, H3C 3J7, Canada}

\author{S.~Pal}
\affiliation{Department of Physics, University of Alberta, Edmonton, T6G 2E1, Canada}

\author{M.-C.~Piro}
\email{mariecci@ualberta.ca}
\affiliation{Department of Physics, University of Alberta, Edmonton, T6G 2E1, Canada}

\author{A.~Plante}
\affiliation{D\'epartement de Physique, Universit\'e de Montr\'eal, Montr\'eal, H3C 3J7, Canada}

\author{S.~Priya}
\affiliation{Materials Research Institute, Penn State, University Park, Pennsylvania 16802, USA}

\author{C.~Rethmeier}
\affiliation{Department of Physics, University of Alberta, Edmonton, T6G 2E1, Canada}

\author{A.~E.~Robinson}
\affiliation{D\'epartement de Physique, Universit\'e de Montr\'eal, Montr\'eal, H3C 3J7, Canada}

\author{J.~Savoie}
\affiliation{D\'epartement de Physique, Universit\'e de Montr\'eal, Montr\'eal, H3C 3J7, Canada}

\author{O.~Scallon}
\affiliation{School of Natural Sciences, Laurentian University, Sudbury, P3E 2C6, Canada}

\author{A.~Sonnenschein}
\affiliation{Fermi National Accelerator Laboratory, Batavia, Illinois 60510, USA}

\author{N.~Starinski}
\affiliation{D\'epartement de Physique, Universit\'e de Montr\'eal, Montr\'eal, H3C 3J7, Canada}

\author{I.~\v{S}tekl}
\affiliation{Institute of Experimental and Applied Physics, Czech Technical University in Prague, Prague, Cz-12800, Czech Republic}

\author{D.~Tiwari}
\affiliation{D\'epartement de Physique, Universit\'e de Montr\'eal, Montr\'eal, H3C 3J7, Canada}

\author{F.~Tardif}
\affiliation{D\'epartement de Physique, Universit\'e de Montr\'eal, Montr\'eal, H3C 3J7, Canada}

\author{E.~V\'azquez-J\'auregui}
\affiliation{Instituto de F\'isica, Universidad Nacional Aut\'onoma de M\'exico, A.P. 20-364, Ciudad de M\'exico 01000, M\'exico}

\author{U.~Wichoski}
\affiliation{School of Natural Sciences, Laurentian University, Sudbury, P3E 2C6, Canada}
\affiliation{SNOLAB, Lively, Ontario, P3Y 1N2, Canada}
\affiliation{Department of Physics, Carleton University, Ottawa, Ontario, K1S 5B6, Canada}

\author{V.~Zacek}
\affiliation{D\'epartement de Physique, Universit\'e de Montr\'eal, Montr\'eal, H3C 3J7, Canada}

\author{J.~Zhang}
\altaffiliation[now at]{ Argonne National Laboratory, 9700 S Cass Ave, Lemont, IL 60439, USA}
\affiliation{Department of Physics and Astronomy, Northwestern University, Evanston, Illinois 60208, USA}

\collaboration{PICO collaboration}
\noaffiliation

\begin{abstract}
The bubble nucleation efficiency of low-energy nuclear recoils in superheated liquids plays a crucial role in interpreting results from direct searches for weakly interacting massive particle (WIMP) dark matter. The PICO collaboration presents the results of the efficiencies for bubble nucleation from carbon and fluorine recoils in superheated C$_3$F$_8$ from calibration data taken with five distinct neutron spectra at various thermodynamic thresholds ranging from 2.1 to 3.9 keV. Instead of assuming any particular functional forms for the nuclear recoil efficiency, a generalized piecewise linear model is proposed with systematic errors included as nuisance parameters to minimize model-introduced uncertainties. A Markov chain Monte Carlo (MCMC) routine is applied to sample the nuclear recoil efficiency for fluorine and carbon at 2.45 and 3.29 keV thermodynamic thresholds simultaneously. The nucleation efficiency for fluorine was found to be $\geq 50 \%$ for nuclear recoils of 3.3 keV (3.7 keV) at a thermodynamic Seitz threshold of 2.45 keV (3.29 keV), and for carbon the efficiency was found to be $\geq 50 \%$ for recoils of 10.6 keV (11.1 keV) at a threshold of 2.45 keV (3.29 keV). Simulated datasets are used to calculate a p value for the fit, confirming that the model used is compatible with the data. The fit paradigm is also assessed for potential systematic biases, which although small, are corrected for. Additional steps are performed to calculate the expected interaction rates of WIMPs in the PICO-60 detector, a requirement for calculating WIMP exclusion limits.

\end{abstract}

\maketitle

\section{\label{S:Intro}Introduction}
Superheated liquids are excellent targets for direct dark matter detection experiments searching for heavy dark matter particles such as weakly interacting massive particles (WIMPs) scattering off atomic nuclei.  The nuclear recoil (NR) that is predicted to result from such a scatter creates a single bubble at the interaction site in the superheated target. World-leading limits have been set on dark matter-nucleus scattering rates based on the observed absence of such bubble nucleation \cite{coupp4, pico2l, pico60_v1, pico60_v2, pico60_v3, picassoref}. In the interpretation of these (and future) experimental results, the bubble nucleation efficiency for NRs as a function of both the thermodynamic state of the chamber and the nuclear recoil energy, is crucial to determining dark matter sensitivity. To characterize this detector response, neutrons are used as proxies for dark matter particles, generating the same NR-induced bubbles as expected from dark matter, but with few-cm mean-free paths and recoil energy spectra that depend on the incident neutron energy.  This paper presents results from the PICO collaboration's campaign over nearly a decade using a variety of neutron sources to calibrate the low-energy (keV-scale) nuclear recoil sensitivity of superheated C$_3$F$_8$.

\subsection{Bubble nucleation by nuclear recoils}
\label{SS: bubble formation}

The thermodynamic basis for bubble nucleation by nuclear recoils is described by Seitz's ``hot-spike'' model \cite{Seitz}, a detailed modern treatment of which is given in \cite{TolmanPICO}, summarized here. This model is based on two well-defined thermodynamic quantities, the critical radius $r_c$ and the energy threshold $Q_{\rm{Seitz}}$. $r_c$ is the radius above which vapor bubbles in the superheated fluid will spontaneously grow, eventually becoming the macroscopic bubbles observed by experiments such as PICO. The critical radius is given by
\begin{equation}
    r_c = \frac{2\sigma}{P_b-P_l},
\end{equation}
where $\sigma$ is the surface tension of the fluid, $P_b$ is the pressure of the vapor filling the bubble, which is approximately equal to the saturation pressure of the fluid at the operating temperature $T$ (controlled experimentally), and $P_l$ is the pressure of the superheated liquid in the bubble chamber (also controlled experimentally). In Seitz's model, nuclear recoils create this critical proto-bubble by locally heating the fluid, and the amount of heat required to create a critically sized proto-bubble is referred to as the Seitz threshold, given by
\begin{equation}\label{eq:Q}
\begin{split}
Q_{\rm{Seitz}} \approx& \text{ } 4\pi r_c^2 \left( \sigma - T \frac{\partial \sigma}{\partial T} \right) + \frac{4\pi}{3}r_c^3 \rho_b (h_b - h_l)\\
& - \frac{4\pi}{3}r_c^3 (P_b - P_l).
\end{split}
\end{equation}
Here, $\rho_b$ is the density of the vapor filling the bubble, and $h_b$ and $h_l$ are the specific enthalpies of the gaseous and liquid states. The three terms in $Q_{\rm{Seitz}}$ represent the energy needed to create the bubble surface, the energy required to vaporize fluid to fill the bubble interior, and the recapture of reversible work present in both of the first two terms.  Any particle interaction injecting heat greater than $Q_{\rm{Seitz}}$ ($\sim$3~keV in the experiments considered here) into a volume small compared to $r_c$ ($\sim$25~nm) will create a critically sized proto-bubble, which will then grow to the macroscopic bubble detected in these experiments.

As a consequence of this idealized model, the efficiency for bubble nucleation would be a step function from 0\% to 100\% when the energy deposited within a critical length scale exceeds the Seitz threshold:
\begin{equation}
\label{E:Edep}
    E_{\rm{dep}} =\int_{0}^{\lambda r_c} \frac{dE}{dx}dx \geq Q_{\rm{Seitz}},
\end{equation}
where $\lambda$ is a unitless scale factor of $O(1)$, often referred to as the ``Harper'' parameter \cite{Harper}.  For low-energy nuclear recoils, this would simplify further to $E_{\rm{dep}}\approx E_r$ (all recoil energy deposited within the critical scale), and a calculation of $Q_{\rm{Seitz}}$ from the temperature and pressure of the superheated target would be sufficient to determine the nuclear recoil detection threshold.  Unfortunately, this is not true for at least three reasons.  First, Eq.~(\ref{eq:Q}) does not include corrections to surface tension at a small radius of curvature, an effect described by the Tolman length~\cite{Tolman} and covered in detail in~\cite{TolmanPICO}.  The Tolman length itself is unknown and leads to $O(0.1)$-keV uncertainties in $Q_{\rm{Seitz}}$.  Second, $Q_{\rm{Seitz}}$ does not account for energy losses that do not contribute to local heating, such as radiative losses (e.g.\ fluorescence), and irreversible work (e.g.\ acoustic radiation), or thermal diffusion transporting heat outside critical radius.  Finally, and most importantly, Eq.~(\ref{E:Edep}) does not reflect event-to-event variation in track structure (i.e.\ straggling in $\frac{dE}{dx}$) which simulations using SRIM \cite{SRIM} show to be significant in nuclear recoils at the energies and length scales considered here.

The effects above can both shift the nuclear recoil detection threshold and, in the case of straggling, broaden it, leading to detection efficiencies below 100\% near the threshold. Indeed, past data with measured monoenergetic recoils \cite{dErrico, dErrico2, Tardif} demonstrate that the nucleation threshold is not an ideal step function. Since there is no quantitative theory to describe these effects, past efforts have invoked \emph{ad hoc} parametrizations for the threshold function, such as an exponential \cite{alphaPICASSO, coupp4, CIRTE}, a sigmoid function \cite{Tardif} and a ``superheated factor'' \cite{Apfel, SIMPLE}. Efforts remain to understand and explain the processes that contribute to the resolution function that shifts and convolutes the Seitz step threshold \cite{Tardif}. Until these effects are fully understood, nucleation efficiencies must be determined by performing dedicated neutron calibrations, as described in this work.

The following sections comprise a global analysis of the PICO collaboration's neutron calibrations to date in superheated C$_3$F$_8$, the target fluid for the PICO-2L, PICO-60, PICO-40L, and PICO-500 dark matter searches.  The general calibration scheme, the specific experimental setups used, and the resulting data and corresponding simulations are described in Sec.\ II.  Section III describes the methods used to extract bubble nucleation efficiencies from this data, including the parametrization of the efficiency function, treatment of systematic uncertainties as nuisance parameters, and the specific Markov chain Monte Carlo (MCMC) technique used to explore the resulting high-dimensional parameter space.  Section IV describes a parametric Monte Carlo study validating the methodology created for this analysis.  Appendices describe the application of this technique to directly constrain WIMP sensitivity in the PICO experiments and present an evaluation of and correction for any bias introduced in this analysis.

\section{\label{S:Calib}Calibration Program -- Data and Simulations}

The objective of the PICO C$_3$F$_8$ nuclear recoil calibration program is to constrain the bubble nucleation efficiency functions $\epsilon_s(E_T,E_r)$, where $\epsilon$ is the probability of bubble nucleation, $s$ indicates the recoil species (carbon or fluorine), $E_T$ is the thermodynamic threshold set by the pressure and temperature of the chamber (equal to $Q_{\rm{Seitz}}$ as calculated in \cite{TolmanPICO}), and $E_r$ is the nuclear recoil energy.  Specifically, we aim to constrain the $E_r$ dependence of these functions at the fixed various $E_T$ employed in the PICO dark matter searches.  Constraints on $\epsilon_s(E_T,E_r)$ come from rate measurements in bubble chambers exposed to known neutron sources.  In general, these constraints take the form of a convolution of the underlying recoil spectrum with the efficiency curve:
\begin{equation}
\label{E:rate}
    R_{\rm{obs}}  = \sum_{s=\mathrm{C,F}}\int_0^\infty R_s(E_r)\cdot\epsilon_s(E_T,E_r)\,dE_r,
\end{equation}
where $R_{\rm{obs}} $ is the experimentally observed rate and the $R_s(E_r)$ are the nuclear recoil spectra (by species) due to the neutron source, determined from simulation.

A single constraint in Eq.~(\ref{E:rate}) is insufficient to deduce the underlying $\epsilon_\mathrm{C}$ and $\epsilon_\mathrm{F}$, as the convolution cannot indicate which recoils are nucleating bubbles -- i.e. threshold shift is indistinguishable from a soft threshold. A suite of measurements with sources producing different recoil spectra, however, is capable of making this distinction. The ability to reconstruct multiple-scattering events in calibration experiments is also key. Each experiment produces not a single rate measurement but a set of measurements giving the one-bubble event rate, two-bubble event rate, etc., with each multiplicity effectively probing a different underlying recoil spectrum. In this sense, Eq.~(\ref{E:rate}) is a simplification applying only to the single-bubble rate in a chamber much smaller than the neutron mean-free path, which is not the case in the measurements presented here. Finally, the $E_T$ dependence of the efficiency curves can be explored by varying the thermodynamic state of a calibration chamber. However, unless assumptions are made relating to the $E_T$ and $E_r$ dependence of the efficiency function, this does little to constrain $E_r$ dependence at the $E_T$ of interest.

\subsection{\label{SS:neutrons}Neutron interactions in C$_3$F$_8$}

\begin{figure}[tb]
  \centering
  \includegraphics[width=0.30\textwidth]{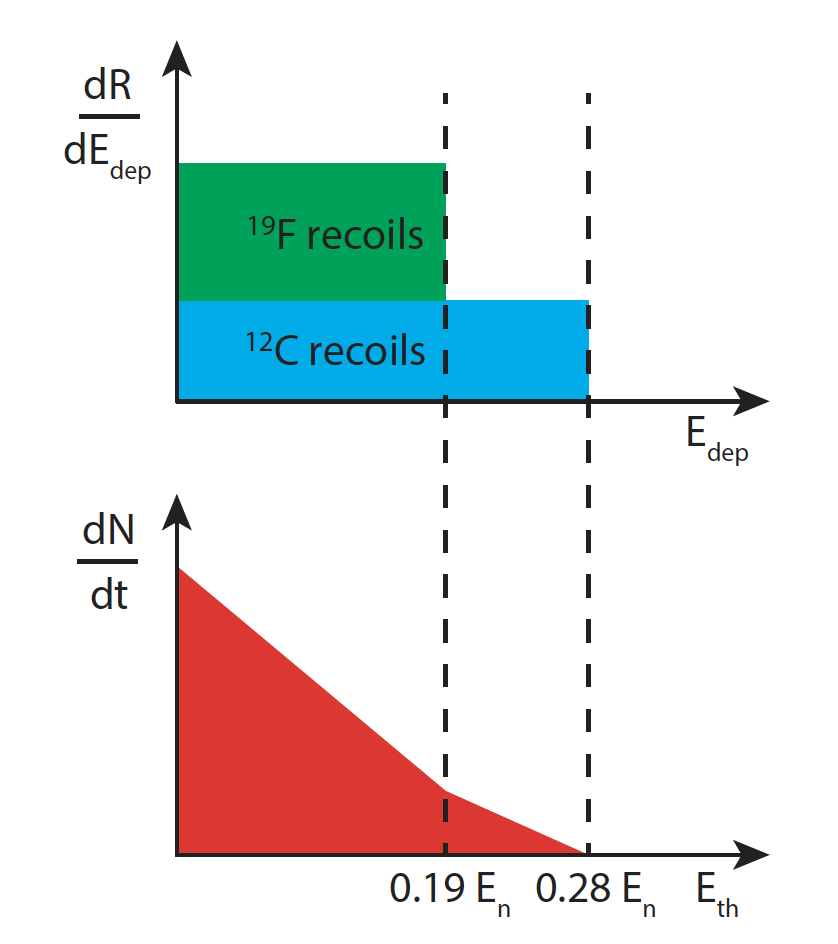}
  \caption{Top: cartoon nuclear recoil energy spectra in the case of isotropic (s-wave) elastic scattering. Bottom: corresponding count rate measured by a threshold detector, as a function of threshold. Copyright Arthur Plante, 2019.}
  \label{EnergySpectrum}
\end{figure}

Neutrons interact predominantly by elastic scattering off of $^{12}$C and $^{19}$F nuclei (with $<1$\% $^{13}$C) in the C$_3$F$_8$ target. The energy spectrum of the recoiling nuclei is described by
\begin{equation}
E_{r} = \frac{2A}{[A+1]^2}(1-\cos\theta)E_{n}
\label{ERecoil}
\end{equation}
where $A$ is the atomic mass of the recoiling nucleus, $\theta$ is the neutron scattering angle in the center of mass frame, and E$_n$ is the incident neutron energy. By fixing $\theta$ = $\pi$, the maximum recoil energies are $E_{r,C_{\mathrm{max}}}$ = 0.28$E_{n}$ and $E_{r,F_{\mathrm{max}}}$ = 0.19$E_{n}$ for carbon and fluorine respectively.  When the incoming neutrons are monoenergetic and the scattering is isotropic, this produces the recoil spectra and rate over threshold shown in Fig.~\ref{EnergySpectrum}.

\begin{figure}[tbh]
  \centering
  \includegraphics[width=0.5\textwidth,trim=20 0 0 50,clip=true]{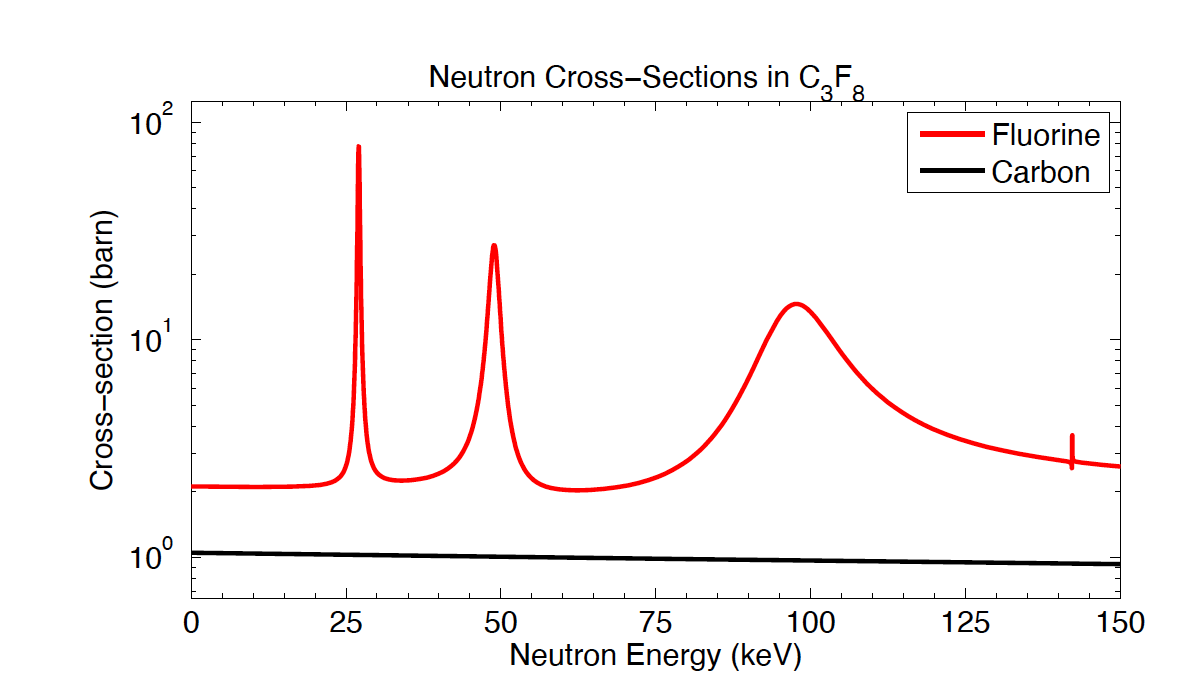}
  \caption{Neutron-carbon and neutron-fluorine elastic scattering cross sections as a function of neutron energy, taken from \protect{\cite{iaea}}, for the neutron energy range used during detector calibration. Copyright Frédéric Girard, 2017.}
  \label{cross-section FC}
\end{figure}

Actual neutron-carbon and neutron-fluorine cross sections are shown in Fig.~\ref{cross-section FC}, illustrating three resonances in the neutron-fluorine scattering cross section in the energy scale of interest.  By calibrating with monoenergetic neutrons at energies on and off these resonances, it is possible to vary the relative rates of carbon and fluorine scattering, allowing separate calibration of $\epsilon_\mathrm{F}$ and $\epsilon_\mathrm{C}$. It should be noted that on-resonance scattering is \emph{not} isotropic, so at these energies, the fluorine recoil spectra deviate from the idealized box-shaped spectrum in Fig.~\ref{EnergySpectrum} \cite{alan_v1}. 

\subsection{\label{SS:Exp}Experimental setup}

\subsubsection{\label{detector} Detectors}

\begin{figure}[tbh]
  \centering
  \includegraphics[width=0.45\textwidth]{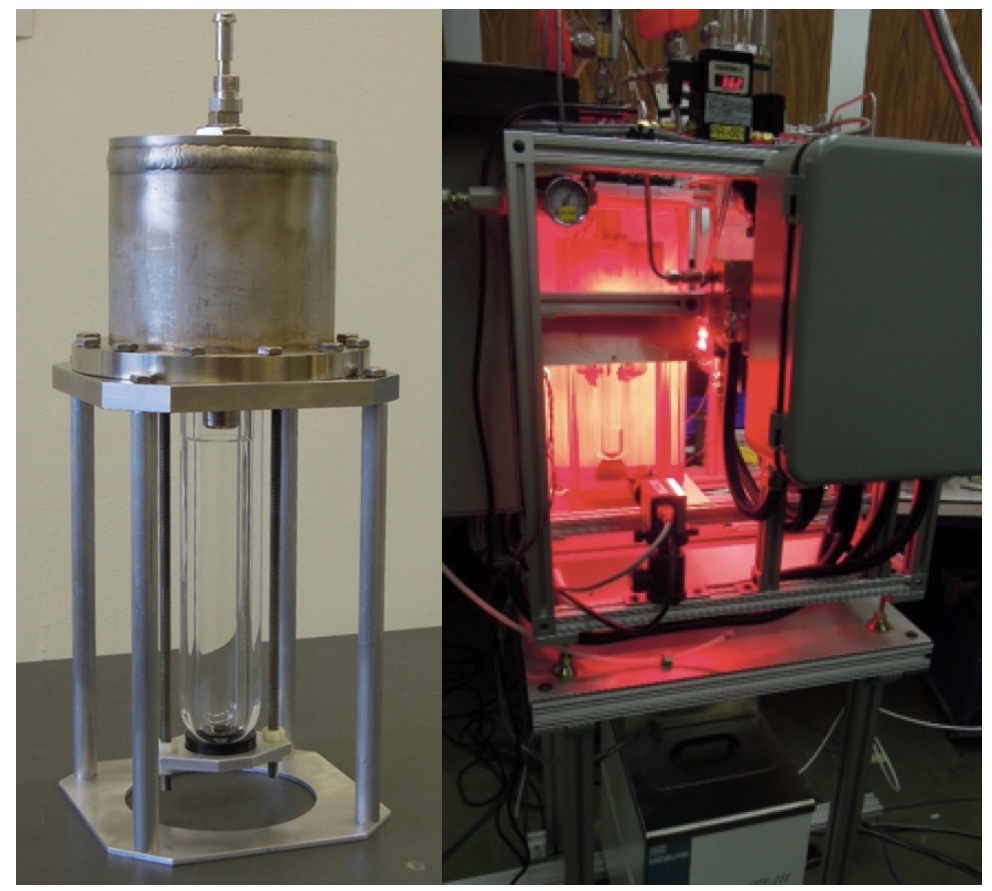}
  \caption{PICO-0.1L detector. Left: the quartz vessel mounted to its hydraulic ``top hat'' containing the bellows used for pressure control of the chamber. Right: the complete setup of the chamber inside its thermal bath equipped with two piezoelectric sensors, two cameras and two LED panels, and CompactRIO-based pressure- and temperature-control system.}
  \label{schematics}
\end{figure}

Two detectors have been employed in the PICO C$_3$F$_8$ nuclear recoil calibration program: the PICO-2L detector at SNOLAB, described in \cite{pico2l}, and a portable version, the PICO-0.1L test bubble chamber, described in \cite{TolmanPICO}. Designed to accomplish calibrations that are difficult in larger chambers, PICO-0.1L consists of a 100-mL, centimeter-thick high-pressure-rated quartz jar attached to a hydraulically driven bellows and immersed in a chilled water bath.  The quartz vessel is filled with 27 $\pm$ 1 gram of C$_3$F$_8$ (19.4 mL of fluid at a density of 1.39 g/mL at 12$^{\circ}$C and 30 psia), as measured by a scale, where the uncertainty is due to losses in the fill lines and the resolution of the scale readout. The remaining space in the vessel is filled with a water buffer, both to create an incompressible volume for pressure control and to avoid any contact between the superheated C$_3$F$_8$ target and the stainless steel surfaces and seals in the bellows system.  Bubble nucleation events are recorded by two piezoelectric acoustic sensors mounted on the jar and by two cameras (150 frames/s) located at 90 degrees from each other. Two LED panels alternate to illuminate the chamber, each in synchronization with one of the cameras. The PICO-0.1L vessel is shown in Fig.~\ref{schematics}, both alone and with the attached imaging and pressure- and temperature-control systems.

\subsubsection{\label{SS:sources}Neutron sources and neutron monitor}
PICO-0.1L was deployed from 2013 to 2019 at the Van de Graaff Tandem accelerator facility at Université de Montréal, where data were taken both with monoenergetic neutron beams produced by the accelerator and with a SbBe photoneutron source.  PICO-2L operated at SNOLAB, with periodic neutron calibrations performed using an AmBe source during the dark matter search.

\paragraph{\label{SS:beam}Monoenergetic neutron beam}

\begin{figure}[tbh]
  \centering
  \includegraphics[width=0.45\textwidth]{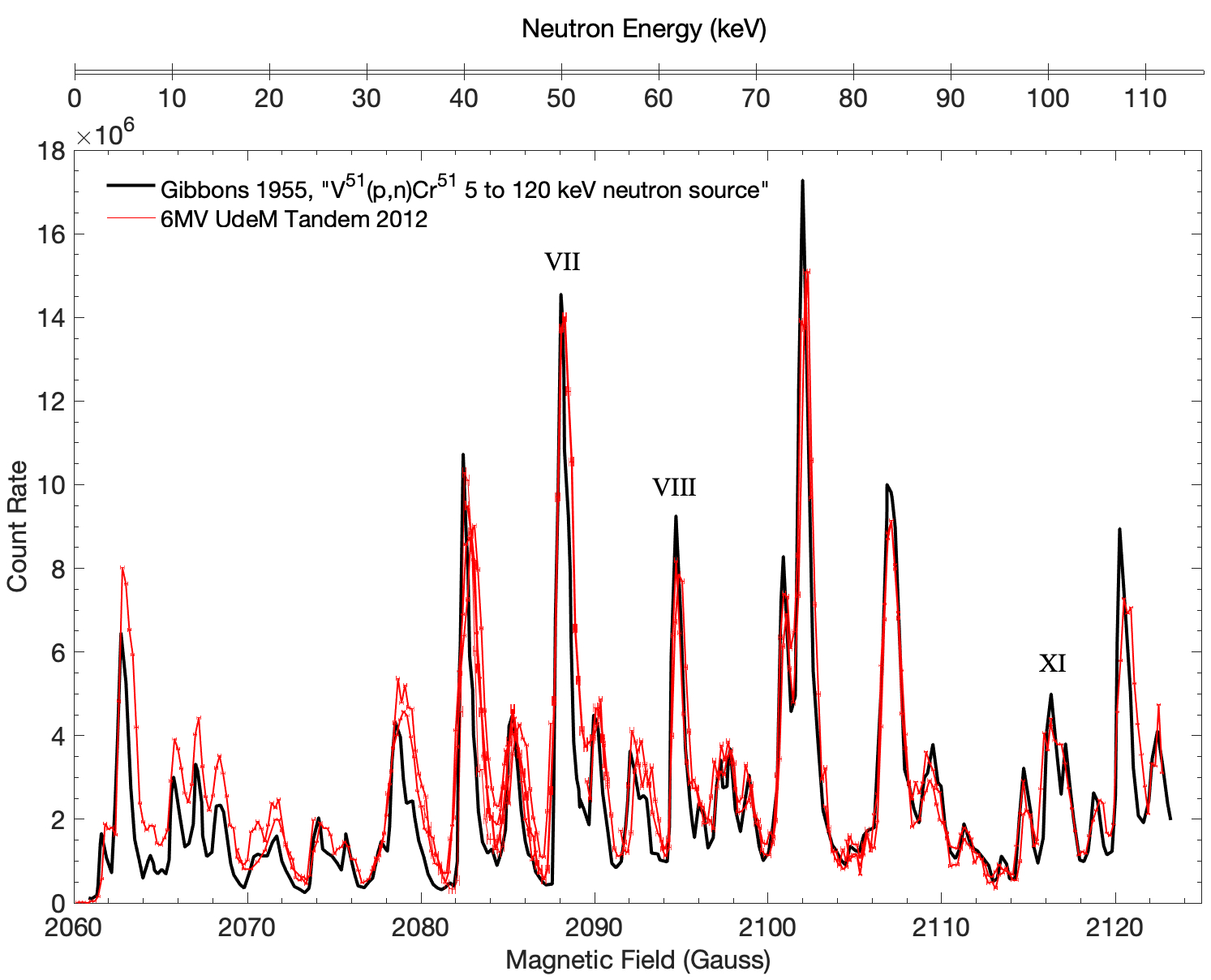}
  \caption{Neutron yield of the $^{51}$V(p,n)$^{51}$Cr reaction as a function of the magnetic field and corresponding neutron energy.  The black curve is taken from \protect{\cite{vanadium}}, and is consistent with measurements done with the Tandem at the Université de Montréal \protect{\cite{LafreniereThesis}} in red. The resonances VII, VIII, and XI chosen for the analysis are also indicated. }
  \label{resonance Vanadium}
\end{figure}

Monoenergetic neutrons were produced with the Tandem Van de Graaff accelerator via the reaction $^{51}$V(p,n)$^{51}$Cr \cite{vanadium, LafreniereThesis}. The energies of the neutrons produced by this reaction are the proton energy minus 1564~keV, the Q value of the reaction so that a monoenergetic proton beam and thin target produce a monoenergetic neutron beam.  The energy resolution of the neutron beam is significantly enhanced when the proton energy is at one of the many sharp resonances shown in Fig.~\ref{resonance Vanadium}. The data considered here were taken at resonances VII, VIII, and XI, corresponding to neutron energies of 50, 61, and 97~keV, respectively. These three $^{51}$V(p,n)$^{51}$Cr resonances were chosen to align on and between the neutron scattering resonances shown in Fig.~\ref{cross-section FC}, with the 50- and 97-keV neutrons on resonance and the 61-keV neutrons off resonance for scattering on fluorine. 

\begin{figure}[tbh]
  \centering
  \includegraphics[width=0.4\textwidth]{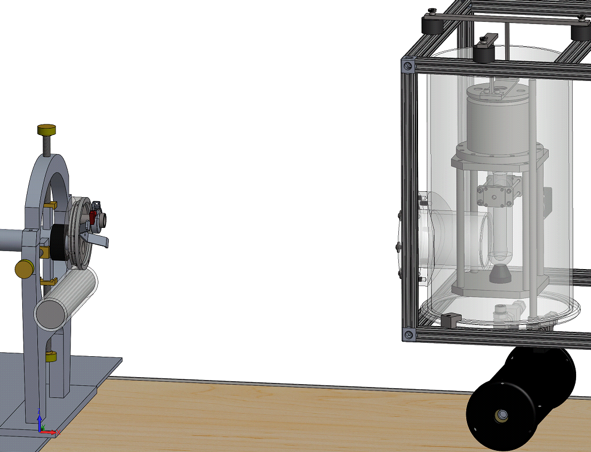}
  \caption{Experimental setup of the test beam calibration at the Université de Montréal. On the left a $^{3}$He counter (horizontal cylinder) is placed underneath the vanadium target located at the end of the proton beam. At the right, the test chamber is facing the beamline with another $^3$He counter located underneath as a second neutron flux monitor. Copyright Miaotianzi Jin, 2019.}
  \label{setup}
\end{figure}

Figure~\ref{setup} illustrates the experimental setup for neutron beam data. The PICO-0.1L test chamber sits directly downstream of the $^{51}$V target, with a penetration through the chilled water bath giving neutrons a direct path to the quartz vessel. The details of this penetration changed between the 2013 and 2014 datasets, as reflected in the analysis in Sec.\ \ref{S: Analysis}. Two $^3$He neutron counters monitor the beam flux, one immediately below the $^{51}$V target and one below the PICO-0.1L water bath. These counters provide the neutron flux normalization described in Sec.~\ref{flux} and provide live feedback to ensure that the proton beam stays at the peak of the selected $^{51}$V(p,n)$^{51}$Cr resonance during data taking.

\paragraph{\label{SS:SbBe}Monoenergetic SbBe photoneutrons}

\begin{figure}[tbh]
  \centering
  \includegraphics[width=0.4\textwidth]{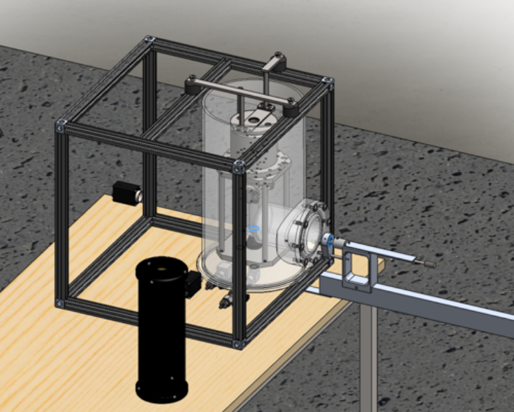}
  \caption{Experimental setup of the SbBe calibration at the Université de Montréal. The PICO-0.1L test chamber is aligned with a graduated source rail to measure the SbBe source position. A $^3$He counter is placed beside the bubble chamber to verify the neutron flux at the chamber. Copyright Miaotianzi Jin, 2019.}  
  \label{setup2}
\end{figure}

To study the detector response at lower recoil energies with the PICO-0.1L test chamber, additional calibration runs were performed using a SbBe photoneutron source, a two-component source that contains a $^{124}$Sb gamma source and a $^{9}$Be conversion target, producing neutrons via the $^9$Be($\gamma$,n) reaction \cite{JCollar,alan2}. There are two significant gamma rays emitted by $^{124}$Sb above the 1665-keV Q value of this reaction: 1691 and 2091~keV, with branching ratios of 48.4\% and 5.7\%, producing 24 and 378 keV neutrons, respectively. A challenge with this calibration is that the source emits $O$(10$^6$) higher gamma flux than neutron flux.  While PICO bubble chambers are in general gamma blind~\cite{TolmanPICO}, the detector does start to become gamma sensitive at Seitz thresholds below 3~keV, increasing the single-bubble event rate. This gamma response is measured by removing the beryllium disk and is considered in the analysis as a background to the SbBe response of the detector. Throughout the measurements, two one-inch thick lead disks were inserted in front of the SbBe source to attenuate the gamma flux, but the gamma-induced single-bubble rate was still deemed too high for useful analysis, and therefore only multibubble events are considered from the SbBe exposure.
The setup for the calibration with the SbBe source (without lead attenuators) is shown in Fig.~\ref{setup2}.

\paragraph{\label{SS:AmBe}AmBe neutrons}

Neutron calibration with an AmBe source was carried out at SNOLAB with the PICO-2L detector filled with C$_3$F$_8$. In contrast to the monoenergetic neutron sources above, this broad-spectrum source of $O$(1)-MeV neutrons generates nuclear recoils far above the detector threshold.  The threshold-constraining power of this measurement comes from the high-statistics, low-background measurement of high-multiplicity events (up to six bubbles and beyond) captured in the two-liter target.  Moreover, the different detector, source, and style of constraint provided by this dataset serve as an important check on the systematic uncertainties associated with each of the neutron calibrations performed. 

\subsection{\label{analysis}Data quality and preparation}
\begin{figure*}[t!]
\center
\includegraphics[width=0.48\textwidth]{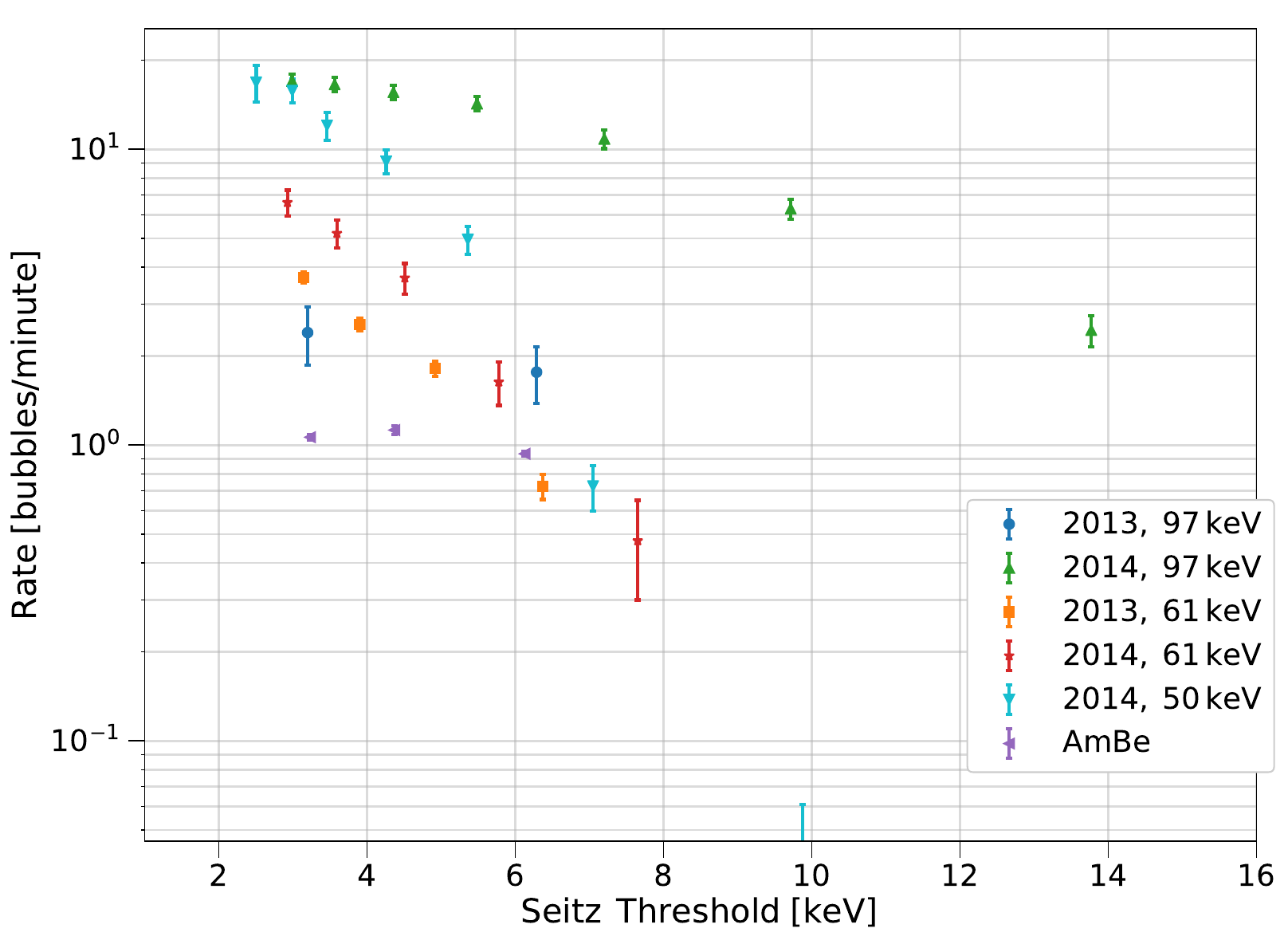}
\includegraphics[width=0.48\textwidth]{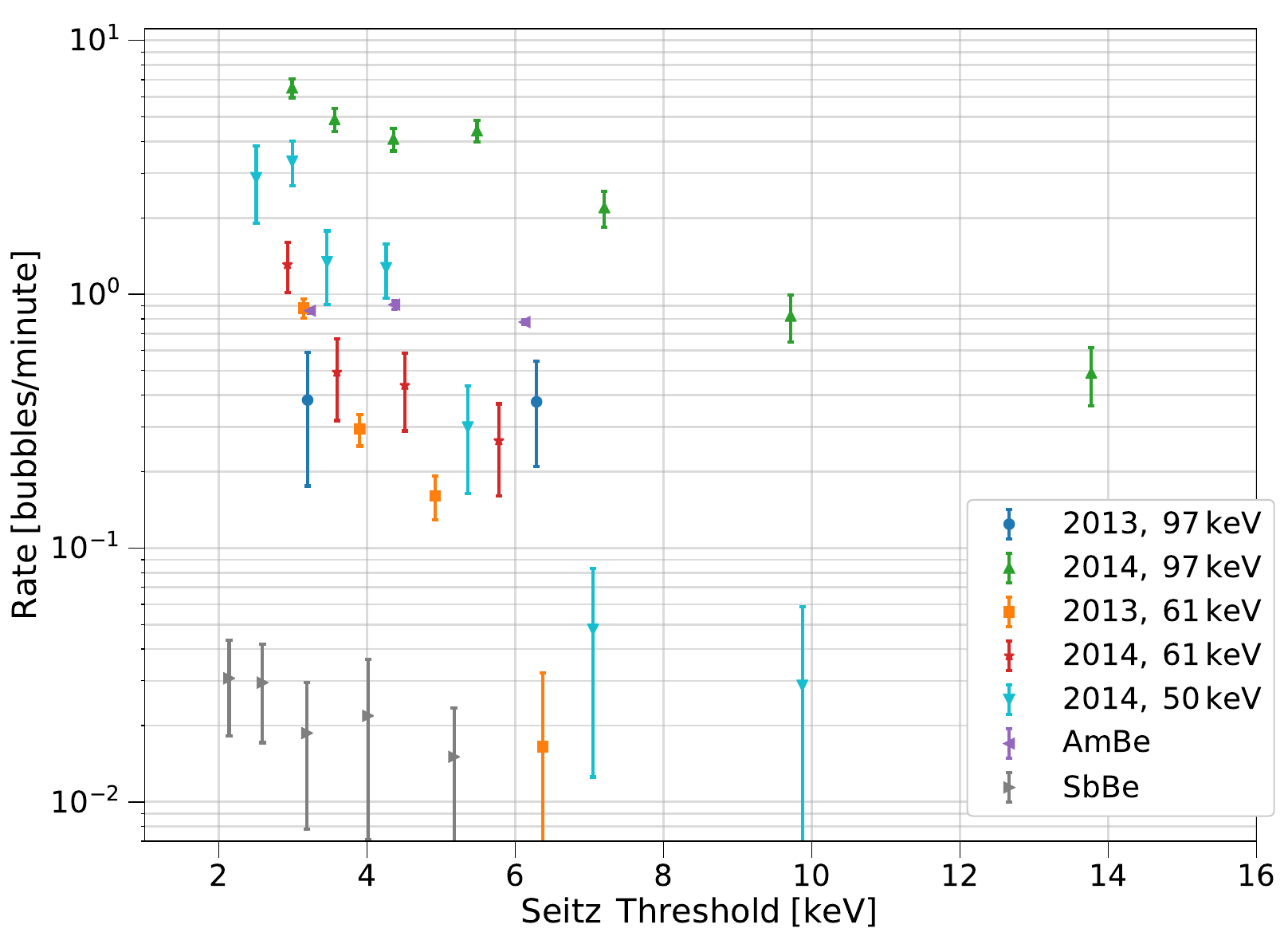}
  \caption{\label{fig:rate_fig} Background-subtracted measured rates of events with one or more bubble (left) and with two or more bubbles (right), in units of bubbles per minute (counting each bubble separately in multibubble events). Error bars show statistical uncertainty only. The background rates subtracted were approximately 0.2 single bubbles/minute for beam data (much less for higher multiplicity events), 0.004 2 bubble events per minute for the SbBe data, and 0 for the AmBe data (below the sensitivity of the background measurement). Data at Seitz thresholds greater than 3.6~keV are not included in the full analysis, but the trends with Seitz threshold illustrate the relative sensitivity of each dataset to the underlying nuclear recoil detection threshold.}
\end{figure*}

Each calibration dataset is reduced to a small set of parameters for subsequent analysis. These parameters include the thermodynamic threshold of the bubble chamber ($Q_{\mathrm{Seitz}}$), total live-time at that threshold, number of events observed (separated by bubble multiplicity), and background bubble rate (also separated by bubble multiplicity) measured without the source/beam in each setup. The background data (dominated by cosmogenic neutrons) was always taken close in time to the corresponding experiment. In the case of the beam experiments, the background rate was found to be consistent over time. For all experiments except for the SbBe data (see paragraph 2b in the previous subsection), it is assumed that gammas from the beam/source produce a negligible background, as the bubble chambers were operated in thermodynamic conditions well beyond the measured electron recoil threshold \cite{TolmanPICO}. The datasets are summarized in Table~\ref{data_org}, and the measured background-subtracted event rates as a function of threshold are shown in Fig.~\ref{fig:rate_fig}.

The above data reduction is performed largely by hand; camera images are visually inspected (hand scanned) to verify bubble counts, a process that is taken to be 100\% efficient.  A series of quality cuts are also applied to eliminate regions in time and/or regions in the detector where chamber performance is compromised.  These cuts include the removal of runs where pressure or temperature control was not functioning as intended; the removal of the first 10 (30) seconds in every PICO-0.1L (PICO-2L) expansion, to allow time for the chamber to reach equilibrium; a fiducial cut in PICO-2L as described in \cite{pico2l}, applied to single-bubble events only; and a fiducial cut in 2013 PICO-0.1L data, removing both single- and mutliple-bubble events where a bubble appears in the top $\sim$cm of the target (which in that run became excessively foamy).  The fiducial cuts in PICO-2L and PICO-0.1L are applied to simulated data as well as to real data.

\begin{table}
\protect\begin{centering}
    \begin{tabular}{| c | c | c | c | c |}
    \hline
    Dataset & Detector & Thresholds & Live-time & Multiplicity \\ 
    & & (keV) & (minutes) & \\ \hline \hline
    97 keV beam & PICO-0.1-2013 & 3.2 & 9.9 & 1,2,3+ \\ \hline
    61 keV beam & PICO-0.1-2013 & 3.1 & 160 & 1,2,3+ \\ \hline
    97 keV beam & PICO-0.1-2014 & 3.0, 3.6 & 21, 20 & 1,2,3+ \\ \hline
    61 keV beam & PICO-0.1-2014 & 2.9, 3.6 & 16, 18 & 1,2,3+ \\ \hline
    50 keV beam & PICO-0.1-2014 & 2.5, 3.0, & 3.1, 7.7, & 1,2,3+ \\ 
    & & 3.5 & 7.3 & \\ \hline
    SbBe & PICO-0.1 & 2.1, 2.6, & 320, 310, & 2,3+ \\
    & & 3.2 & 300 & \\ \hline
    AmBe & PICO-2L & 3.2 & 2200 & 1,2,3,4, \\
    & & & & 5,6,7+ \\  \hline
    \end{tabular}
\par\end{centering}
\protect\caption{Datasets included in this analysis. Each row corresponds to a single combination of detector and neutron source. One or more thermodynamic thresholds were employed for each setup (listed in keV), with varying live-times (in minutes). Different bubble multiplicities are considered for each setup, with the `+' indicating that the final multiplicity bin includes events with higher bubble multiplicity as well.}
\label{data_org}
\end{table}

\subsection{\label{SS:simu}Simulations of neutron scattering}

For each neutron calibration dataset, a GEANT4 \cite{Geant4} (SbBe and AmBe data) and/or MCNP-POLIMI \cite{MCNP} (beam and AmBe data) Monte Carlo simulation is performed to provide the recoil energy spectra for single- and multiple-scatter events within the detector. These are used, in conjunction with hypothesized bubble nucleation efficiency functions, to calculate the expected event rate by bubble multiplicity for each calibration experiment. Neutron scattering cross sections for both transport through the geometry and signal production in the detector are taken from Evaluated Neutron Data File ENDF/B-VII \cite{iaea} for both MCNP and GEANT4 \cite{G4NDL}, with corrections from \cite{alan_v1} for the differential scattering cross sections at the fluorine resonances, which ENDF incorrectly treats as s-wave (isotropic).

\begin{figure*}[t!]
\center
\includegraphics[width=0.48\textwidth]{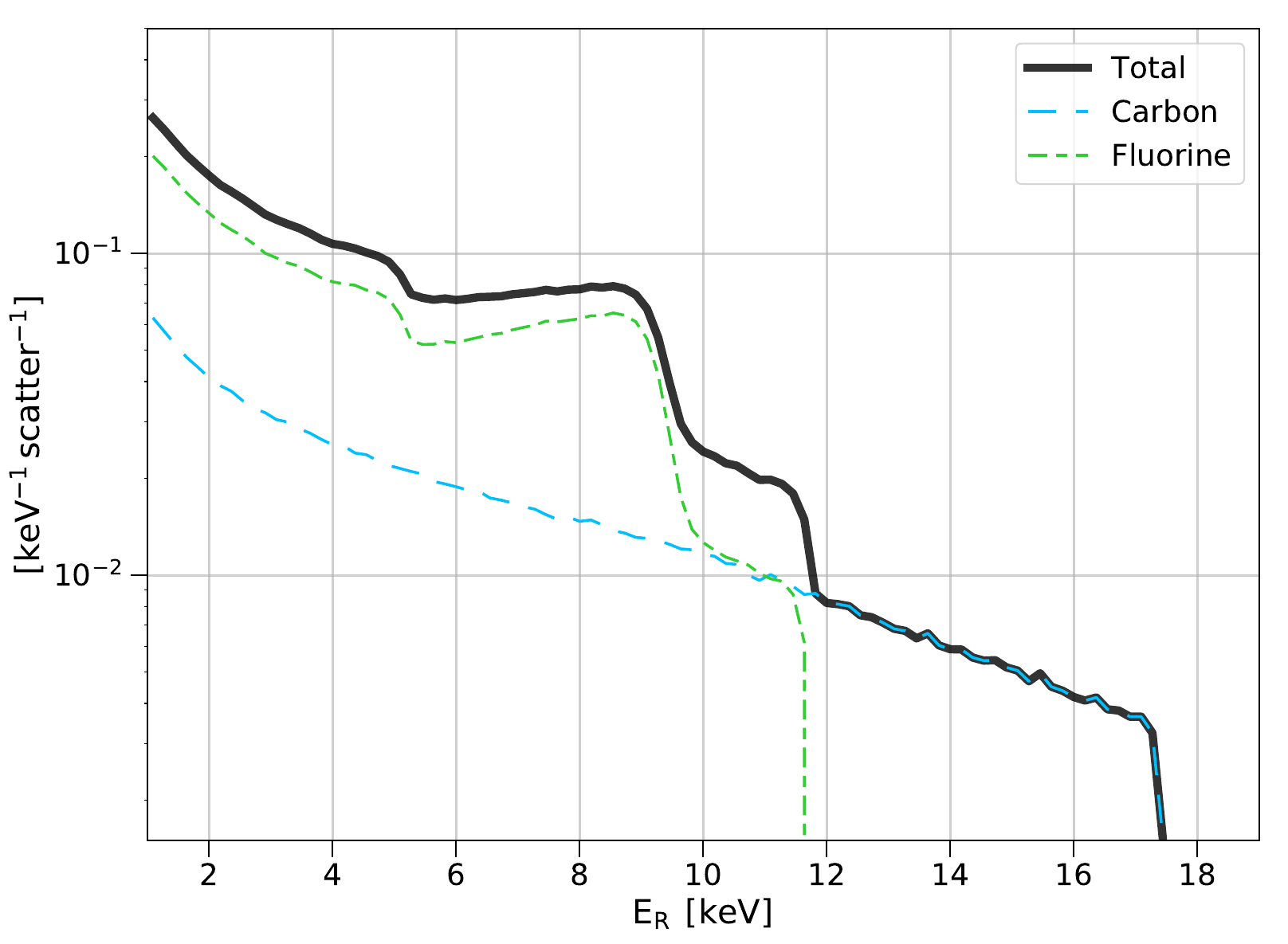}
\includegraphics[width=0.48\textwidth]{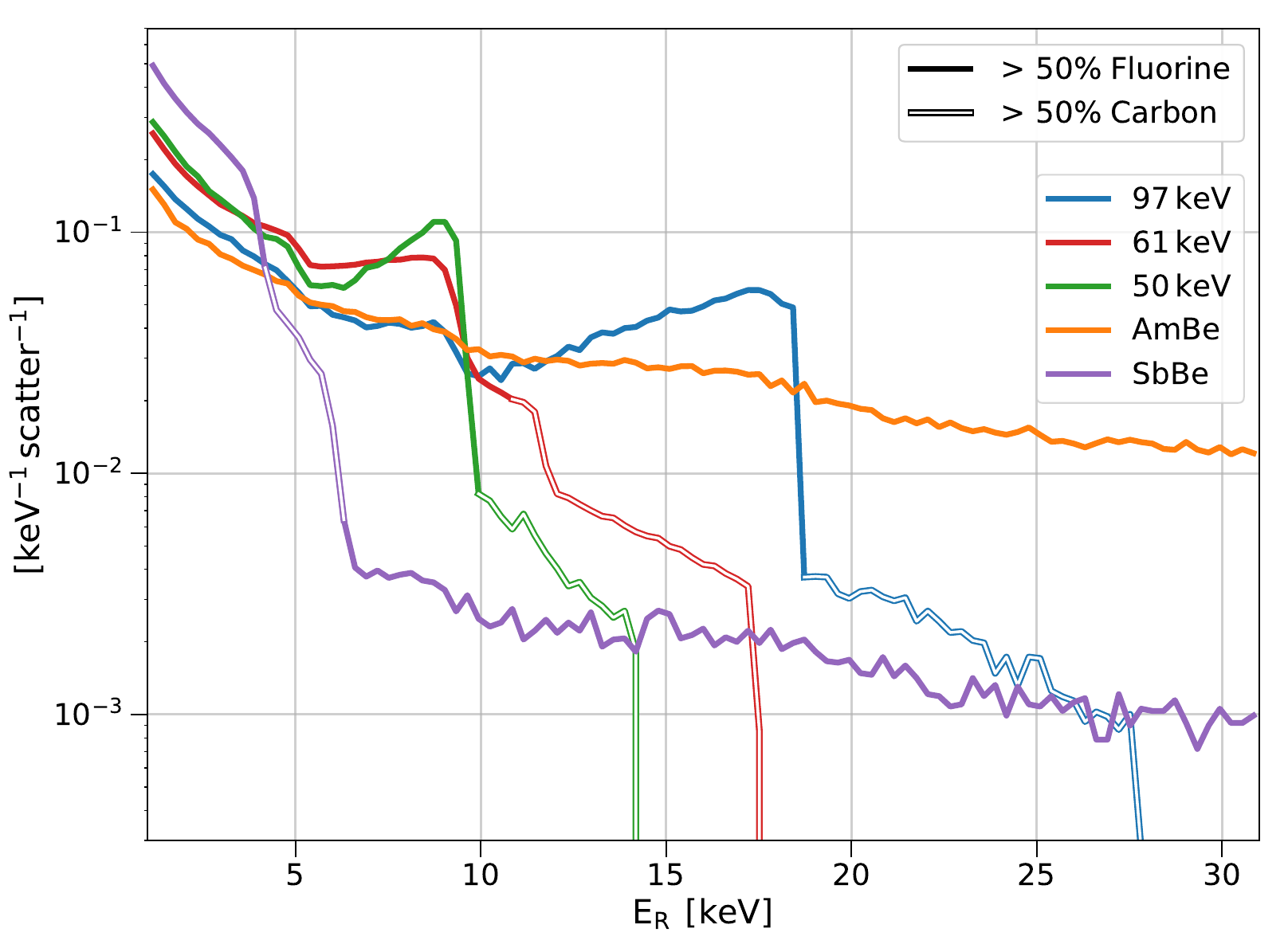}
  \caption{\label{fig:sim_spec} Left: simulated recoil energy spectrum for the 61 keV beam experiment (2013) showing the separate contributions of scatters off of fluorine and carbon. Right: simulated recoil energy spectra for the 97 keV (2013), 61 keV (2013), and 50 keV (2014) beam experiments, as well as the SbBe and AmBe experiments. The dominance of scattering off of fluorine or carbon atoms is indicated by solid or hollow lines respectively. The spectra in the left and right panels are normalized to the total number of scatters in each simulation off of either target species.}
\end{figure*}

Each of the three neutron sources considered requires a slightly different treatment for particle generation.  For the beam simulations, the energy-angle relation for neutrons coming from the $^{51}$V(p,n)$^{51}$Cr reaction is included directly in the MCNP-POLIMI neutron source definition, based on data in \cite{vanadium}.  A similar energy-angle relation exists for the SbBe photoneutron source, but because the Sb gammas are emitted isotropically into a large Be target, a simple neutron source definition capturing this relation is not possible.  Furthermore, this relation cannot be simulated natively in MCNP-POLIMI -- while MCNP-POLIMI does simulate the $^9$Be($\gamma$,n) reaction, it does so by treating the incoming photon as a zero-momentum particle, and therefore misses the angular dependence of the resulting neutron energy.
To address this shortcoming, a photon-only simulation of the Sb source and Be target was run and neutron production was inserted manually to generate a list of neutrons with a given initial position, direction, and energy.  This list was then used as the input for a neutron simulation in GEANT4. The AmBe exposure in PICO-2L was simulated with both MCNP-POLIMI and GEANT4 to test different source spectra included in the respective simulation packages. No significant discrepancies in recoil spectra were found, and the MCNP-POLIMI simulation was selected for this analysis. Examples of the resulting simulated recoil energy spectra are shown in Fig.~\ref{fig:sim_spec}, which shows the general dominance of fluorine (carbon) scatters at low (high) recoil energies, as was depicted in Fig.~\ref{EnergySpectrum}.

\subsection{\label{flux}Neutron flux measurements}
Constraints from both beam and SbBe data in PICO-0.1L are more powerful when the neutron flux (as well as energy) is well known. This is not true for the PICO-2L AmBe data, where constraints derive largely from ratios of rates at different multiplicities. This section describes the ancillary measurements used to anchor the neutron flux from these sources.

\subsubsection{Neutron beam flux\label{SS:beamflux}}
The relative flux from the neutron beam is continuously monitored by the $^3$He counter sitting immediately below the $^{51}$V target.  Translating this to an absolute neutron flux at the PICO-0.1L chamber is challenging in simulation, requiring precise knowledge of the materials and geometry surrounding the $^{51}$V target and $^3$He counter. It is also difficult to verify via \emph{in situ} measurement -- the second $^3$He counter near the chamber is insufficient for this purpose, due primarily to uncertainties in the composition and geometry of materials near that counter, including the PICO-0.1L water bath and support structure.  For this reason, two independent measurements were performed to verify the absolute neutron beam flux. 

\begin{figure}[tbh]
  \centering
  \includegraphics[width=0.45\textwidth]{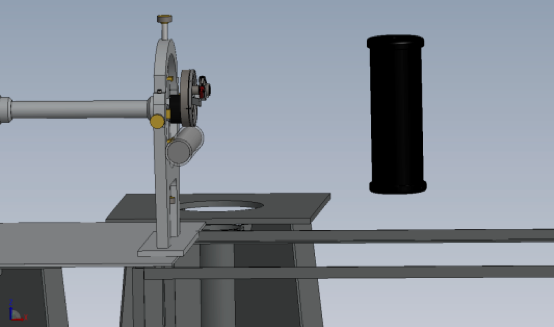}
  \caption{Experimental setup at the Université de Montréal for measuring the neutron flux. Two $^{3}$He counters were used: one directly below the $^{51}$V target and one suspended directly downstream of the target. Copyright Miaotianzi Jin, 2019.}
  \label{setup3}
\end{figure}

\paragraph{$^3$He-only measurement}
The first flux measurement was performed with the two $^{3}$He counters only, with the PICO-0.1L chamber and all unnecessary structures removed. The target-side $^3$He counter remained in its usual position below the $^{51}$V target, and the chamber-side counter was suspended downstream of the target as shown in Fig.~\ref{setup3}. All geometry details, including the beam pipe, were entered into an MCNP \cite{MCNP} Monte Carlo to simulate the neutron capture rates in the two counters. The comparison between the measured neutron capture ratio of both counters with Monte Carlo simulations at each beam energy is shown in Table \ref{hanging_he3}. At all three energies, the agreement is reasonable, and while there is a trend to slightly overpredict the near target to downstream ratio, the source of this bias is unclear and no correction is attempted. Instead, these uncertainties are included as nuisance parameters in Sec.\ \ref{SS:systematics}.

\begin{figure}
  \centering
  \includegraphics[width=0.5\textwidth]{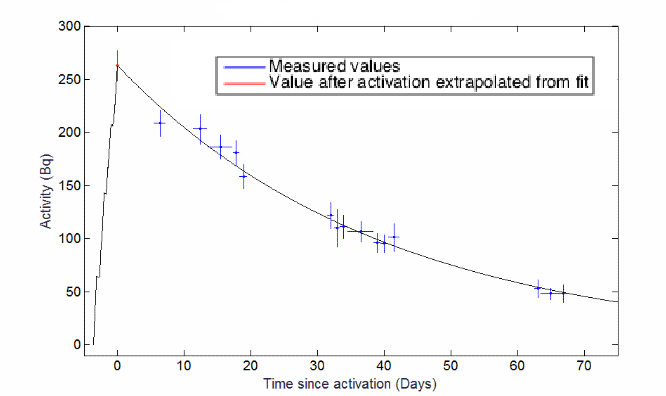}
  \caption{$^{51}$Cr activity in the $^{51}$V target after exposure to the proton beam.  These data were taken at a proton beam energy corresponding to 50-keV neutron production, with a fresh $^{51}$V target (no $^{51}$Cr activity prior to exposure).  A single-parameter fit with fixed $^{51}$Cr half-life gives the activation rate of the target when exposed to the beam.}
  \label{activation}
\end{figure}

\begin{table}
\begin{center}
    \begin{tabular}{| c | c | c | c |}
    \hline
    Energy & Measured ratio & Simulated ratio & Measurement / simulation \\ \hline
    50~keV & 2.28 $\pm$ 0.07 & 2.26 $\pm$ 0.08 & 1.01 $\pm$ 0.05 \\ \hline
    61~keV & 2.02 $\pm$ 0.07 & 2.26 $\pm$ 0.08 & 0.89 $\pm$ 0.04 \\ \hline
    97~keV & 2.07 $\pm$ 0.10 & 2.21 $\pm$ 0.07 & 0.93 $\pm$ 0.05 \\ \hline
    \end{tabular}
    \protect\caption{Measured neutron capture rate ratio of the two $^{3}$He counters in the setup shown in Fig.~\ref{setup3} (near-target rate / downstream rate), compared to Monte Carlo simulation.  Uncertainties in the measurement are statistical uncertainties from the number of captures recorded  in the two $^3$He counters.  Uncertainties on the simulated ratio are highly correlated between the different energies and include uncertainties on the $^3$He content of the counters and the density/makeup of the neutron moderator surrounding the counters.}
    \label{hanging_he3}
\end{center}
\end{table}
\paragraph{Target activation ($^{51}$Cr) measurement}
The second flux calibration technique is to measure the $^{51}$Cr activity in the $^{51}$V target before and after exposure to the proton beam, directly measuring the number of neutrons produced via the $^{51}$V(p,n)$^{51}$Cr reaction.  The 320-keV gammas produced by the $^{51}$Cr electron-capture decay are measured with a Ge-detector, calibrated using a $^{133}$Ba source (303- and 356-keV gammas) with precisely known activity ($\pm$ 3\%) and geometry similar to the $^{51}$V target disk. An exponential decay fit on the gamma activity was performed with a fixed 27.7-day half-life to determine the total quantity of $^{51}$Cr produced (see Fig.~\ref{activation}).   
Table \ref{activationtable} presents the results obtained at 50-keV neutron energy (the only beam energy where this measurement was performed).  This check is not directly used elsewhere in the analysis, but the agreement seen in Table \ref{activationtable} validates the use of the target-side $^3$He counter and corresponding MCNP simulation to fix the absolute neutron beam flux.

\begin{table}[tbh]
\begin{center}
    \begin{tabular}{|c|c|}
%    \multicolumn{2}{c}{Neutrons produced during activation run at 50 keV} \\
    \hline
    Calculation method & Neutrons produced ($\times 10^8$) \\
    \hline
    $^3$He plus MCNP simulation & 9.18 $\pm$ 0.52 \\ 
    \hline
    $^{51}$Cr activity & 9.52 $\pm$ 0.51 \\
    \hline
    \hline
    Ratio & 1.04 $\pm$ 0.08\\
    \hline
    \end{tabular}
    \protect\caption{Total neutrons produced during a sample 50-keV neutron beam run, measured via the $^3$He counter (with MCNP simulation to convert the $^3$He capture rate to neutron production rate) and via $^{51}$Cr activity in the $^{51}$V target.}
    \label{activationtable}
\end{center}
\end{table}

\subsubsection{SbBe neutron flux}
Rather than base the neutron flux from the SbBe source on the underlying $^{124}$Sb activity and simulated $^9$Be($\gamma$,n) rate, both of which contain significant uncertainty, the neutron flux from the SbBe source was measured directly, using the same $^3$He counter that was used as the downstream counter for the beam neutron flux normalization.  Multiple measurements with the SbBe source and $^3$He counter in various relative positions and orientations give a total neutron yield of 209 $\pm$ 22 neutrons per second (corrected for the $^{124}$Sb half-life in individual datasets), where the uncertainty reflects the measurement-to-measurement variation in the SbBe neutron yield.

\section{\label{S: Analysis}Constraints on bubble nucleation efficiency for nuclear recoils}

The objective of this analysis is to constrain $\epsilon_\mathrm{C}(E_T,E_r)$ and $\epsilon_\mathrm{F}(E_T,E_r)$ -- the efficiency curves for nucleation with carbon and fluorine respectively - at $E_T$ of 2.4 and 3.29~keV, two operating conditions of the PICO-60 dark matter detector \cite{pico60_v3}.  This is accomplished through a maximum likelihood analysis, treating each rate measurement as an independent Poisson variable.  We express this likelihood as
\begin{align}
\label{E-LL}
\log \mathcal{L} = & \sum_{i}\sum_{j}\left[-\nu_{i,j}(\{x_{s,p,T}\},\{s_k\})+k_{i,j}+ \nonumber {\color{white} \frac{|}{|}} \right. \\ & \left. k_{i,j}\log\left(\frac{\nu_{i,j}(\{x_{s,p,T}\},\{s_k\})}{k_{i,j}}\right)\right] -  \sum_{k}\frac{s_k^2}{2},
\end{align}
\noindent where $\nu_{i,j}$ are the expected numbers of events (including background rate) with bubble multiplicity $j$ for experiment $i$, and $k_{i,j}$ are the corresponding observed number of events. One ``experiment'' refers to a combination of detector, threshold, and neutron source (see Table~\ref{data_org}, which includes the bubble multiplicities considered for each experiment).  The $\nu_{i,j}$ depend both on the hypothesized efficiency curves, parametrized by \{${x_{s,p,T}}$\}, and on a set of nuisance parameters \{$s_k$\}, each of which represents the number of standard deviations that a given source of systematic uncertainty deviates from its nominal value.  The total likelihood is frequently cast as an effective chi-square statistic, defined as 
\begin{equation}
\chi^2 = -2 \times \log \mathcal{L}.
\label{chi2}
\end{equation}

Section~\ref{SS:model} describes the adopted parametrization of the efficiency curves, defining the $x_{s,p,T}$'s. Section~\ref{SS:systematics} defines nuisance parameters considered and the magnitudes of the corresponding systematic uncertainties, and the analysis results are given in Sec.\ \ref{SS:bestfit}. Appendix A describes the Markov chain Monte Carlo (MCMC) technique for exploring the resulting 34-dimensional parameter space. Due to the large number of variables and computationally expensive likelihood function, a regular MCMC was deemed to be too inefficient to use. Therefore a novel MCMC algorithm was developed to force more efficient exploration of the boundary of the likelihood function to produce the fit results shown in Sec.\ \ref{SS:bestfit}.

\subsection{\label{SS:model} Parametrization of the efficiency functions}

\begin{figure}[tb]

  \centering
  \includegraphics[width=0.47\textwidth]{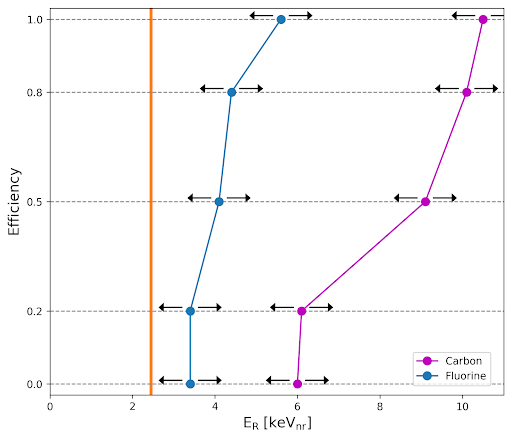}
  \caption{\label{F:piecewiselinear} Cartoon showing the piecewise linear model used in this analysis. Each blue (purple) dot represents a movable set point of fluorine (carbon) at efficiencies of 0, 0.2, 0.5, 0.8, and 1.  In this analysis, the carbon threshold is required to lie to the right of the fluorine threshold, which in turn must lie to the right of the Seitz threshold (orange vertical line).}  
\end{figure}

To avoid artificially constraining the shape of the bubble nucleation efficiency curve, the curve is modeled as a piecewise-linear function, as shown in Fig.~\ref{F:piecewiselinear}. Each ``knot'' in the function is held at fixed nucleation probability but allowed to translate in recoil energy.  By increasing the number of knots in the function, any shape efficiency curve can be approximated (at the cost of a higher-dimensional parameter space to explore).
The $\{x_{s,p,T}\}$ are the knot locations on the recoil energy axis for a set of these piecewise-linear functions, where the index $s$ indicates recoil species ($s\in$\{C, F\}), the index $p$ indicates nucleation probability ($p\in$\{0, 0.2, 0.5, 0.8, 1\}), and the index $T$ indicates the Seitz threshold ($T\in$\{2.45~keV, 3.29~keV\}), giving a total of 20 free parameters setting the efficiency curves in this analysis.

Several physics-driven constraints are imposed on the $\{x_{s,p,T}\}$.  First, the efficiency curves are taken to be monotonic in both recoil energy and threshold:
\begin{eqnarray}
    \frac{\partial \epsilon_s}{\partial E_r}\ge0& \quad \rightarrow \quad & x_{s,p_{i+1},T} \ge x_{s,p_i,T} ,\\
    \frac{\partial \epsilon_s}{\partial E_T}\le0& \rightarrow& x_{s,p,T_{i+1}} \ge x_{s,p,T_i} .
\end{eqnarray}
Second, the bubble nucleation efficiency for carbon recoils is assumed to be lower than the bubble nucleation efficiency for fluorine recoils at the same energy:
\begin{equation}
       \epsilon_\mathrm{C}(E_T,E_r) \le \epsilon_\mathrm{F}(E_T,E_r)\quad\rightarrow\quad x_{\mathrm{C},p,T} \ge x_{\mathrm{F},p,T}.
\end{equation}
This constraint is based on nuclear recoil stopping models and on simulations using SRIM~\cite{SRIM} of nuclear recoil cascades in C$_3$F$_8$, both of which indicate that carbon recoils deposit their energy over a much greater distance than fluorine recoils of the same energy.  Finally, no bubble nucleation is allowed for recoils with energies below the Seitz threshold:
\begin{equation}
x_{s,0,T} \ge T.
\end{equation}

The majority of the datasets used in this analysis (see Table~\ref{data_org}) were taken at thresholds near but not precisely at the thresholds used in PICO-60, so some $E_T$ dependence must be assumed when using calibration data to constrain the $x_{s,p,T}$.  This is done by extrapolating from the nearest-neighbor threshold fencepost.  That is, for calibration data taken at threshold $E_T$, the efficiency curves applied to the calibration data are given by
\begin{equation}
    x_{s,p}(E_T) = \frac{E_T}{\widehat{T}}x_{s,p,\widehat{T}},
\end{equation}
where $\widehat{T}$ is the nearest fencepost (2.45 or 3.29~keV) to $E_T$.  This is done to diminish the influence of high-threshold calibration data on the low-threshold efficiency curve, and vice versa.

\subsection{\label{SS:systematics}Systematic uncertainties}
Two types of systematic uncertainty are considered in this analysis:  uncertainties on the thresholds at which calibration data are taken, and uncertainties on the neutron exposure in each dataset.  In both cases, uncertainties are treated as multipliers to the nominal threshold or neutron exposure, where the multiplier is sampled from a log-normal distribution with an average of one.  Each source of uncertainty effects some subset of the data used in this analysis, though the magnitude of a given systematic effect may vary between datasets (e.g.\ a miscalibrated pressure transducer is a single source of systematic uncertainty, but has a larger impact at high threshold than it does at low threshold).

\subsubsection{\label{SSS:thresholds}Uncertainties on calibration thresholds}
The five nuisance parameters capturing systematic uncertainty on the thresholds at which calibration data are taken are shown in Table~\ref{nuisance_1}. The four pressure and temperature (PT) uncertainties represent the errors caused by miscalibration of a chamber's pressure and temperature transducers (typically taken to be 1 psi and 0.1$^{\circ}$C), as well as variations in pressure and temperature during the course of a run. These variations are modeled as coherent across a given experimental setup to minimize the number of nuisance parameters included in the analysis.  Treating the PT variations in this way is conservative, in that it decreases the $\chi^2$ cost associated with coherent fluctuations in the threshold.

The final nuisance parameter impacting threshold is a global parameter representing uncertainty translating thresholds in our calibration chambers to thresholds in PICO-60.  Put another way, the two threshold fenceposts given in Sec.\ \ref{SS:model} are \emph{defined} to be the PICO-60 operating condition, and this nuisance parameter captures all uncertainty on the corresponding threshold energy.  This includes fundamental uncertainties in the Seitz threshold calculation (e.g.\ uncertainty in the Tolman length, see Sec.\ \ref{SS: bubble formation} \cite{Tolman,TolmanPICO}) as well as systematic miscalibration of the pressure and temperature transducers in PICO-60.  

\begin{table}
\protect\begin{centering}
\begin{tabular}{|c|c|}
    \hline
Nuisance  & \% uncertainty \\
parameter & (multiple entries indicate $E_T$ dependence) \\ \hline
PT uncertainty: & 97 keV:  8.0 \\
beam 2013 & 61 keV:  8.0 \\\hline
PT uncertainty: & 97 keV:  7.0 / 9.9  \\
 beam 2014& 61 keV:  1.7 / 2.5  \\
 & 50 keV:  7.0 / 14  \\\hline
PT uncertainty: SbBe & 6.5 / 7.0 / 7.5 \\\hline
PT uncertainty: Ambe & 8.0  \\\hline
Fencepost thresholds & 3.0  \\\hline
\end{tabular}\protect
\par\end{centering}
\protect\caption{Five nuisance parameters describing uncertainty in calibration thresholds, and their one-sigma amplitudes.  Each of the four calibration setups has an associated nuisance parameter capturing systematic pressure/temperature uncertainty.  The final row captures uncertainty in the location of the two reference threshold fenceposts, which are defined by the PICO-60 operating conditions.}    
\label{nuisance_1}
\end{table}

\subsubsection{Uncertainties on neutron exposure}

Exposure uncertainties come in two parts: geometric uncertainties impacting neutron transport to the bubble chamber target, and uncertainties on the source strength.  For SbBe and AmBe data, these uncertainties are combined into a single nuisance parameter, but for neutron beam data they are separated, allowing for a common geometry uncertainty in each setup with separate source-strength nuisance parameters at each beam energy (see Table~\ref{nuisance_2}).

The dominant contribution for geometric uncertainty in PICO-0.1L comes from the length of the path through the water bath (i.e.\ the few-mm gap between the end of the neutron conduit and the quartz vessel) that neutrons must traverse to reach the target fluid. The width of this water gap in the 2013 setup is measured to be 2 $\pm$ 1 mm using camera images of the chamber.  The water bath and neutron conduit were rebuilt in 2014 with significantly reduced uncertainty on the gap distance, resulting in the exposure uncertainties listed in line 1 and line 2 in Table \ref{nuisance_2}.  Geometric uncertainties in PICO-0.1L in the SbBe setup are negligible relative to the source strength uncertainty. Uncertainties on source strength, both for the neutron beam flux and the SbBe photoneutron source, are given by the source strength measurements described in Sec.\ \ref{flux}. Geometric uncertainty in PICO-2L is estimated based on the different neutron flux seen at the target fluid in the GEANT4 and MCNP simulations described in Sec.\ \ref{SS:simu}.  This uncertainty is much larger than uncertainty on the AmBe source strength but has little impact on the final analysis as this nuisance parameter turns out to be well constrained by the global calibration dataset.

\begin{table}
\protect\begin{centering}
\begin{tabular}{|c|c|}
\hline
Nuisance parameter  & \% uncertainty \\ \hline
PICO-0.1L geometry 2013 & 7.5  \\ \hline
PICO-0.1L geometry 2014  & 3.1 \\\hline
Source: Beam, 2013, 97 keV  & 8.3  \\\hline
Source: Beam, 2013, 61 keV & 4.7  \\\hline
Source: Beam, 2014, 97 keV & 5.9  \\\hline
Source: Beam, 2014, 61 keV & 5.0  \\\hline
Source: Beam, 2014, 50 keV & 5.4  \\\hline
Source + geometry: SbBe & 10.3  \\\hline
Source + geometry: AmBe & 26  \\\hline
\end{tabular}\protect
\par\end{centering}
\protect\caption{Nine nuisance parameters describing uncertainties in neutron exposure and their one-sigma amplitudes.  The first two rows describe uncertainties in chamber geometry, impacting all 2013 and 2014 neutron beam data, respectively.}    
\label{nuisance_2}
\end{table}

\begin{figure}[t]
  \centering
  \includegraphics[width=0.49\textwidth]{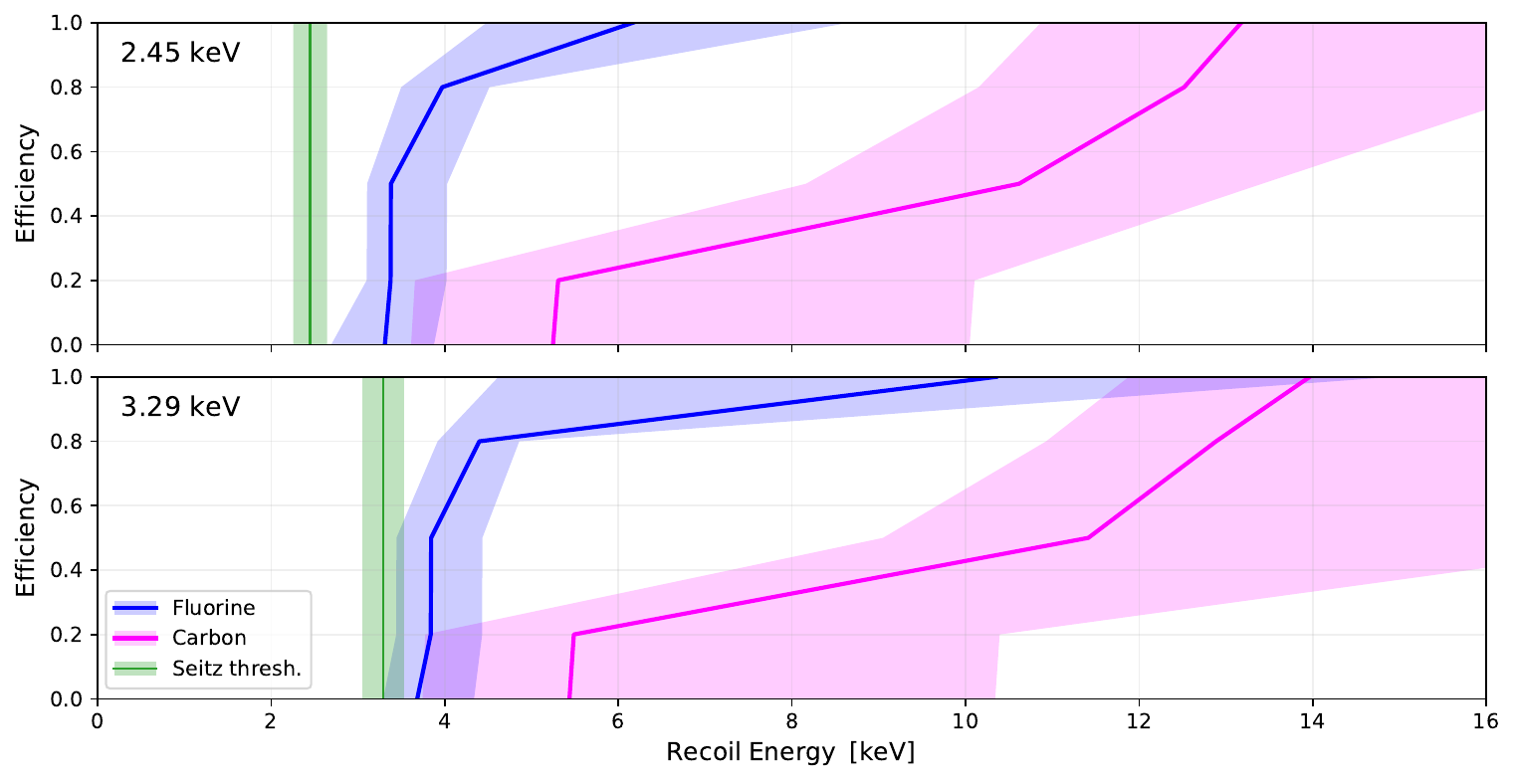}
  \caption{\label{fig:efficiency} Best-fit and 1$\sigma$ error bands for the nucleation efficiency curves of fluorine (blue) and carbon (magenta), at the thresholds used in the PICO-60 WIMP search~\protect{\cite{pico60_v3}}. Error bands indicate the 1$\sigma$ range of each knot position, so that every efficiency curve within 1$\sigma$ of the best fit falls in the shaded area (but not every curve falling in the shaded area is within 1$\sigma$ of the best fit).  The corresponding Seitz thresholds are shown as a vertical green line, with the green band indicating the systematic threshold uncertainties described in Table~\ref{nuisance_1}.}
\end{figure}

\begin{figure}[h!]
\centering
\includegraphics[width=0.44\textwidth]{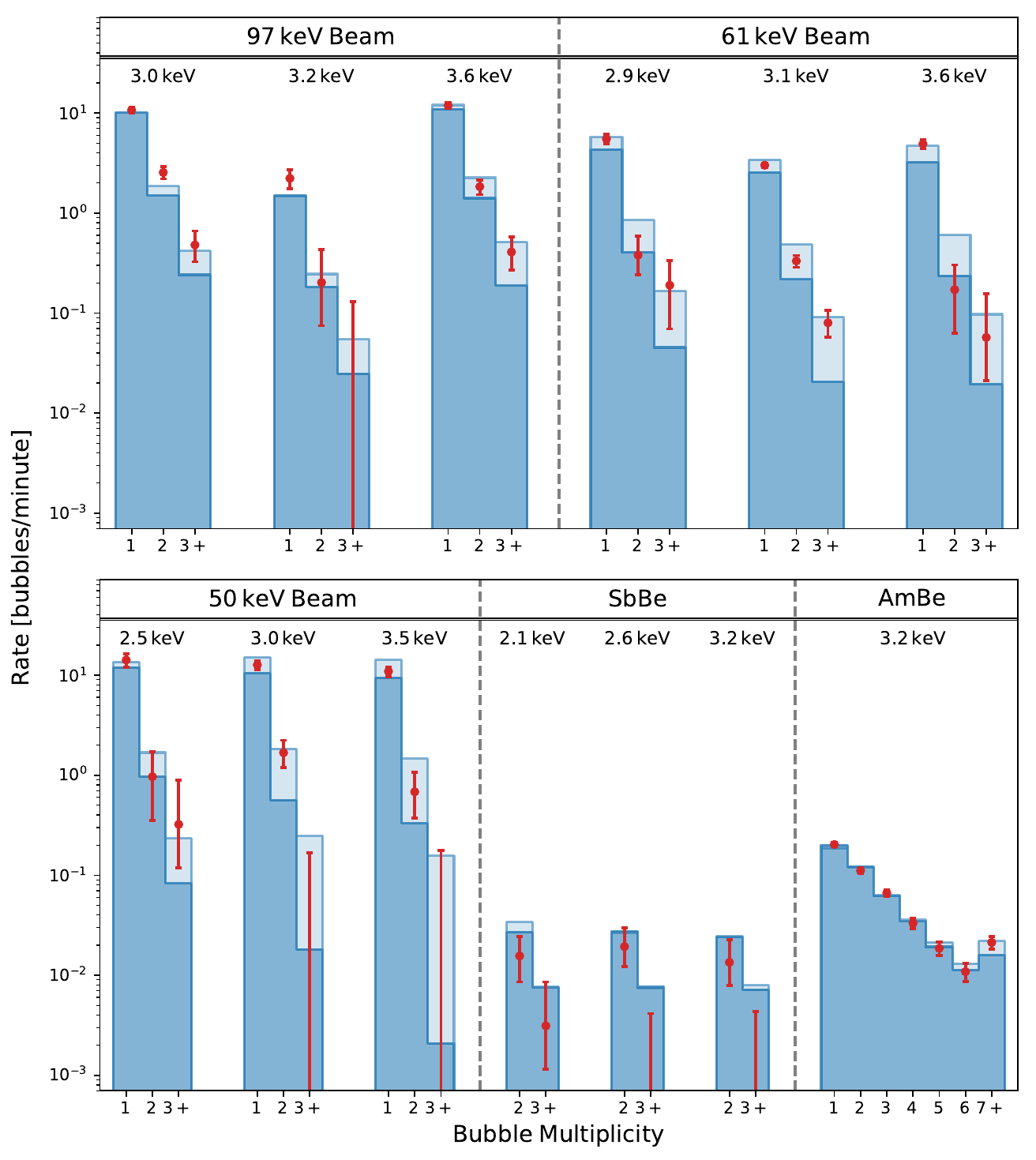}
\caption{\label{multiplicity} Comparison between the measured bubble rate (in bubbles per minutes) and fit of the data. The red points are the experimental data points with Feldman-Cousins $1 \sigma$ confidence intervals as error bars on the Poisson mean in each bin \cite{feldman}. The blue histograms are the result of the MCNP/GEANT4 simulations convolved with the bubble efficiency curves fit to the data; the empty blue bars represent the $\pm$1$\sigma$ error window of the fit.}
\end{figure}

\subsection{Results\label{SS:bestfit}}
\begin{figure}[h]
  \centering
  \includegraphics[width=0.46\textwidth]{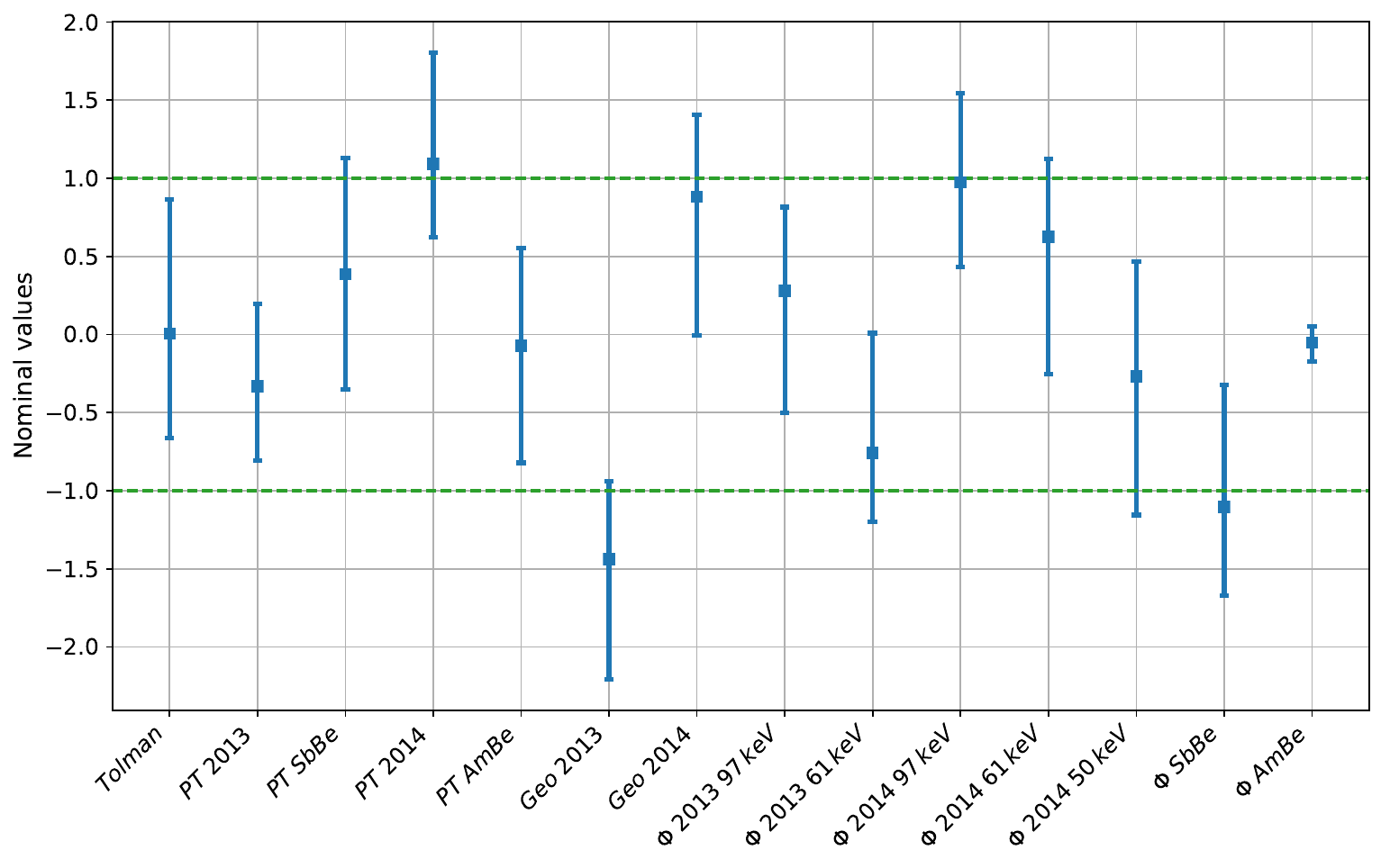}
  \caption{\label{fig:nuisance} Posterior constraints on nuisance parameters. Each nuisance parameter on this plot has a Gaussian prior with mean of $0$ and standard deviation of $1$.  11 of 14 best-fit values fall within the 1$\sigma$ prior band, and in all cases the 1$\sigma$ posterior error bars overlap with the 1$\sigma$ prior band.  The rightmost point, representing the neutron flux from the AmBe source in PICO-2L, is highly constrained by the data, largely due to the high-bubble-multiplicity resolved in that dataset.}
\end{figure}

The model was fit to the data using an iterative Markov chain Monte Carlo approach to perform the maximum likelihood fit. This novel method allowed for efficient exploration of the high-dimensional likelihood function, and is described in Appendix A. The best-fit nucleation efficiency curves and their 1$\sigma$ error bands are shown in Fig.~\ref{fig:efficiency}. The error bands were approximated to be the bounds of the parameter space where $\log \mathcal{L} \geq \mathrm{max} \left \{ \log\mathcal{L}\right \} - \frac{1}{2}$ (implicitly profiling over all other parameters) for convenient visualization. These results are consistent with those used in \cite{pico60_v2,pico60_v3}. A comparison between these results and the data can be seen in Fig.~\ref{multiplicity} (showing predicted vs observed bubble rates in calibration data) and Fig.~\ref{fig:nuisance} (showing posterior constraints on nuisance parameters). Notably, while the fluorine curve turns on near the Seitz threshold at both fenceposts, all nucleation efficiency curves deviate significantly from the Seitz threshold at high nucleation probability and have nontrivial shapes that are not readily comparable to standard functional forms such as a sigmoid or exponential function. It is also worth noting that the carbon bubble nucleation threshold is significantly higher than the fluorine threshold, as expected -- the by-hand constraint enforcing that relation has almost no impact on the result.  The fluorine efficiency curves do tend to have smaller uncertainty bands than carbon, primarily because a greater proportion of the bubbles produced in the calibration  experiments, particularly at low recoil energy, are created from recoils with fluorine.

\section{Parametric Monte Carlo Study}
\label{S:MCstudy}

\begin{figure*}[tbh]
\center
\includegraphics[width=0.48\textwidth]{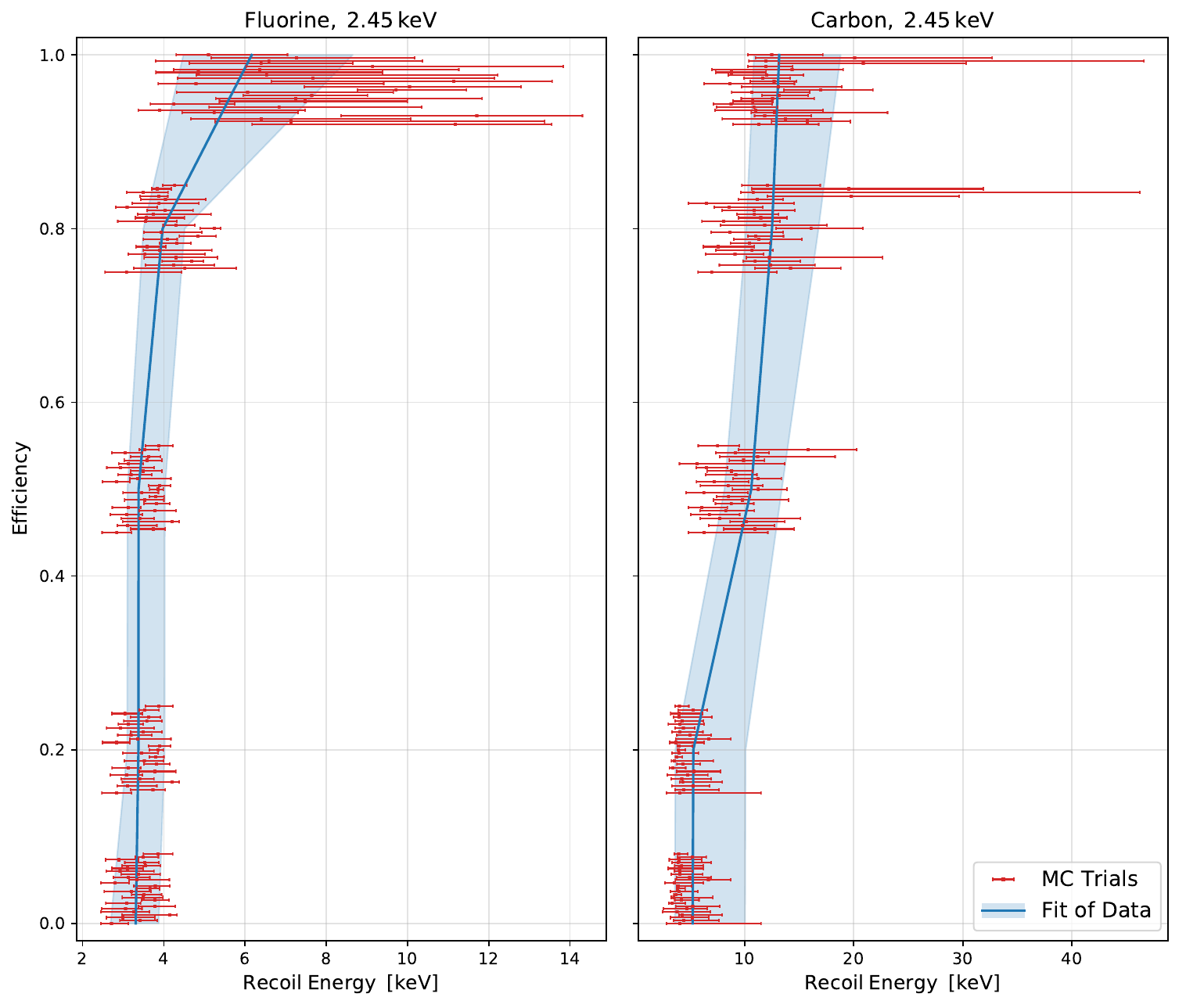}
\includegraphics[width=0.48\textwidth]{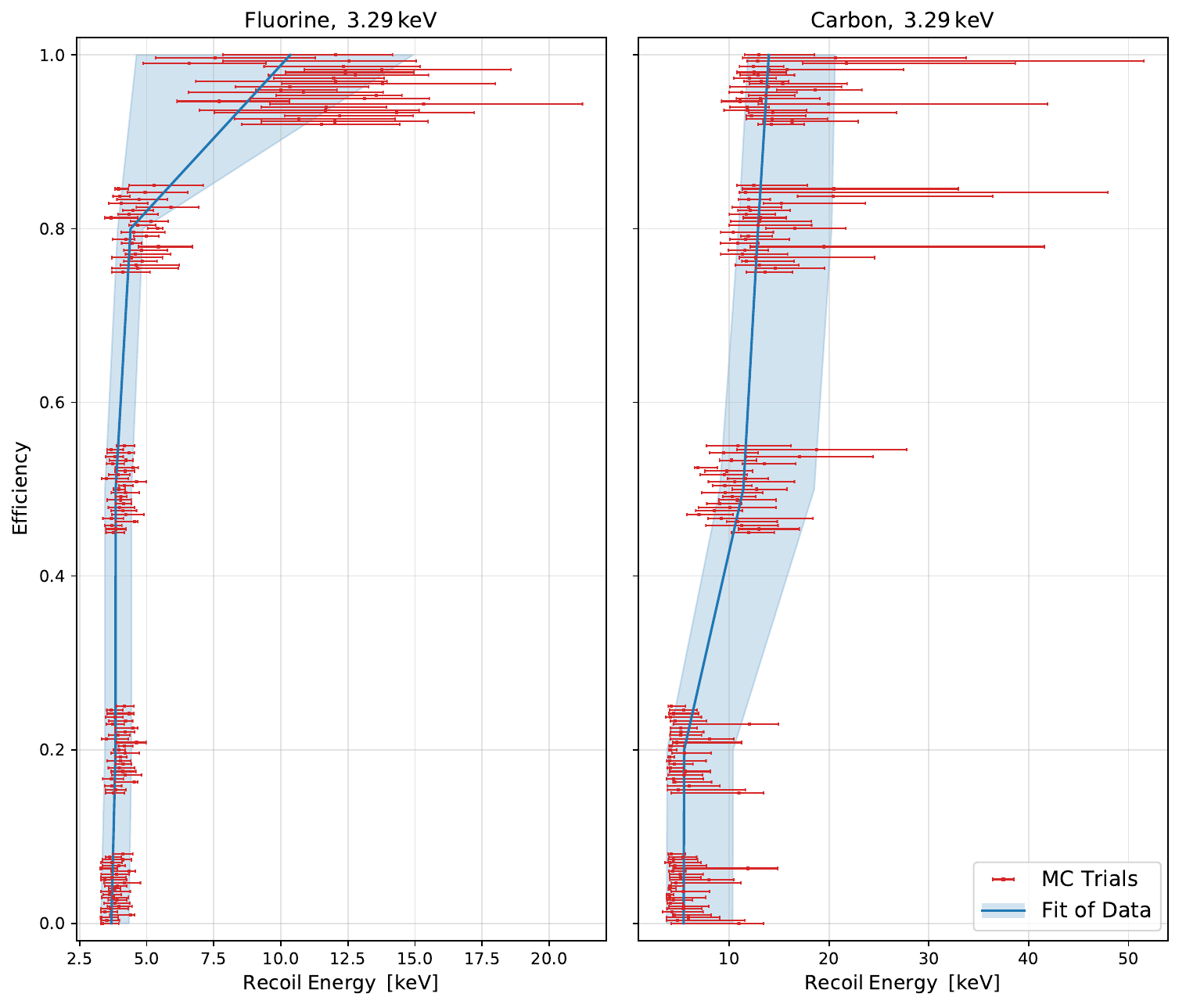}
  \caption{\label{fig:bias} Fits of $25$ simulated datasets (red) compared to the best fit and $1 \sigma$ band of the actual data (light blue) for the thresholds of 2.45 keV (left) and 3.29 keV (right).}
\end{figure*}

A useful test of this methodology that can be performed is to generate and fit simulated datasets. Doing so makes it possible to check if the fitting procedure used produces unbiased results (see Appendix A), validate the method used to fit the model to the data, and enable interpretation of the final posterior $\chi^2$ value obtained for the fit of the data. To produce the simulated data, the best-fit model depicted in Fig.~\ref{fig:efficiency} is used (hence ``parametric'') to generate randomly drawn count data. These simulated datasets are then fit with the same procedure described in Appendix A. The results of these $25$ fits compared to the original best fit are shown in Fig.~\ref{fig:bias}.

Because the nucleation efficiency analysis presented in this paper is \emph{ad hoc} and untested on other data, it is prudent to wonder if the likelihood function describing the data is unbiased. The parametrically simulated data described in this section can be used to assess this by comparing the fit results to the input values. One can see in Fig.~\ref{fig:bias} that in some cases there are small, but persistent systematic offsets between the true value of a parameter and the $25$ fits, such as at an efficiency of $0$ for carbon with a threshold of 3.29 keV.  However, when considering the bias in each parameter estimate as a function of all parameters (not necessarily assuming a constant bias), the various biases tend to have opposing effects. Consequently, the efficiency curves corrected for the aggregate bias - shown in Fig.\ \ref{fig:NFB} (Appendix B) - are only very slightly different than the original result shown in Fig. \ref{fig:efficiency}. Full details of this model biases analysis are presented in Appendix B.

\begin{figure}[tbh]
  \centering
  \includegraphics[width=0.43\textwidth]{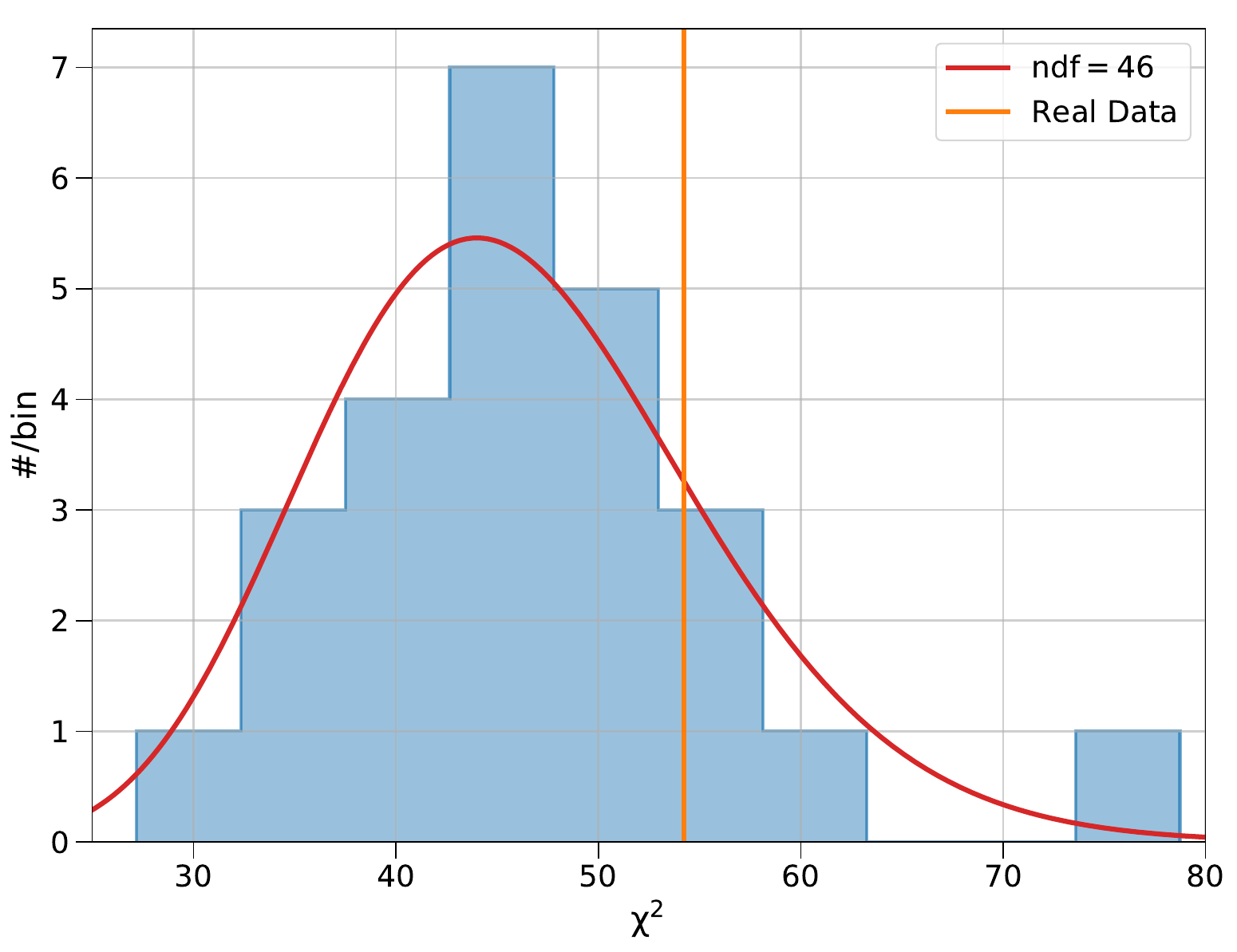}

  \caption{\label{fig:chi2} $\chi^2$ data for the $25$ simulated datasets (blue histogram), the best-fit $\chi^2$ distribution to that data (red curve), and the value of $\chi^2$ obtained for the real data (orange line).  Based on the best-fit $\chi^2$ distribution, the real data give a p-value of 0.19.}
\end{figure}

Another powerful result of this study is the ability to interpret the posterior $\chi^2$ values obtained as proper goodness-of-fit statistics. This is nontrivial due to the strong correlations between many of the fit parameters, increasing the effective number of degrees of freedom far beyond the nominal expectation of $\mathrm{\#\, points - \#\, parameters = 17}$. The distribution of final $\chi^2$ values for the simulated datasets is shown in Fig.~\ref{fig:chi2}. By fitting this distribution with the $\chi^2$ probability density function, an effective number of degrees of freedom of $46$ is obtained. In that case, the value of $\chi^2$ obtained for the real data of $54.2$ is reasonable, being $< 1 \sigma$ away from the expectation, with a p value of $p = 0.19$.

\section{\label{S:summary}Summary}

A set of monoenergetic and broad-spectrum neutron sources has been used to calibrate the nuclear recoil response of C$_3$F$_8$ in PICO detectors. A flexible, many-parameter functional form for nucleation efficiency is assumed for data-driven analysis. A modified MCMC approach was used for the exploration of the high-dimensional likelihood space to simultaneously estimate nucleation efficiency at 3.29 and 2.45-keV thermodynamic thresholds, for both carbon and fluorine recoils. All of these nucleation efficiency curves deviate from their corresponding Seitz threshold. The $50 \%$ nucleation efficiency point for fluorine recoils was found to be 3.3 keV (3.7 keV) at a thermodynamic Seitz threshold of 2.45 keV (3.29 keV), and for carbon, the efficiency was found to be $50 \%$ for recoils of 10.6 keV (11.1 keV) at a threshold of 2.45 keV (3.29 keV). These results are in agreement with earlier analyses of this data published by the PICO collaboration. The fact that the fluorine efficiency curves at both thresholds are nearly the same suggests that there is not enough data at the lower threshold of 2.45 keV to fully constrain this result. Also, the relative lack of calibration data for carbon alone resulted in wider uncertainty bands for those efficiency curves.

Beyond these initial results, an extensive study of simulated datasets was carried out. This validated the convergence criteria used for the MCMC exploration of the real calibration data. Additionally, the $\chi^2$ results for the simulated datasets were used to calculate a p value for the fit of the real data of $0.19$, confirming that the model is adequately flexible to describe the data and that the MCMC procedure resulted in a reasonable fit. Finally, these results were also used to characterize small systematic biases in this fit paradigm, providing bias-corrected nucleation efficiency curves. This correction is small and therefore does not invalidate previous results obtained with this data.

\section*{ACKNOWLEDGEMENTS}
The PICO collaboration wishes to thank SNOLAB and its staff for support through underground space, logistical and technical services. SNOLAB operations are supported by the Canada Foundation for Innovation and the Province of Ontario Ministry of Research and Innovation, with underground access provided by Vale at the Creighton mine site. We wish to acknowledge the support of the Natural Sciences and Engineering Research Council of Canada (NSERC) and the Canada Foundation for Innovation (CFI) for funding, and the Arthur B. McDonald Canadian Astroparticle Physics Research Institute. We acknowledge that this work is supported by the National Science Foundation (NSF) (Grants No.\ 0919526, No.\ 1506337, No.\ 1242637, No.\ 1205987, and No.\ 1828609), by the U.S.\ Department of Energy (DOE) Office of Science, Office of High Energy Physics (Grants No.\ DE-SC0017815 and No.\ DE-SC0012161), by the DOE Office of Science Graduate Student Research (SCGSR) award, by the Department of Atomic Energy (DAE), Government of India, under the Centre for AstroParticle Physics II project (CAPP-II) at the Saha Institute of Nuclear Physics (SINP), and Institutional support of Institute of Experimental and Applied Physics, Czech Technical University in Prague   (IEAP CTU, DKRVO). This work is also supported by the German-Mexican research collaboration Grants No.\ SP 778/4-1 (DFG) and No.\ 278017 (CONACYT), Project No.\ CONACYT CB-2017-2018/A1-S-8960, Dirección General de Asuntos del Personal Académico, Universidad Nacional Autónoma de México (DGAPA UNAM) Grant No.\ PAPIIT-IN108020, and Fundación Marcos Moshinsky. This work is partially supported by the Kavli Institute for Cosmological Physics at the University of Chicago through NSF grants 1125897 and 1806722, and an endowment from the Kavli Foundation and its founder Fred Kavli. We also wish to acknowledge the support from Fermi National Accelerator Laboratory under Contract No.\:DE-AC02-07CH11359, and from Pacific Northwest National Laboratory, which is operated by Battelle for the U.S.\ Department of Energy under Contract No.\:DE-AC05-76RL01830. We also thank the Digital Research Alliance of Canada \cite{computecanada} and the Centre for Advanced Computing, ACENET, Calcul Qu\'ebec, Compute Ontario, and WestGrid for computational support. The work of M.~Bressler is supported by the Department of Energy Office of Science Graduate Instrumentation Research Award (GIRA). The work of D.~Durnford is supported by the NSERC Canada Graduate Scholarships -- Doctoral program (CGSD).

\section*{\label{A:MCMC method}Appendix A: MCMC algorithm}

Constraining the nuclear recoil bubble nucleation efficiency curves is achieved by mapping the log-likelihood given by Eq.~(\ref{E-LL}) over the 34-dimensional parameter space using the \emph{emcee} MCMC library \cite{emcee}, a Python MCMC toolkit designed especially for high-dimensional spaces.  This tool employs multiple interdependent ``walkers'' to explore the parameter space, evaluating the likelihood at multiple points in each step in the Markov chain and creating a proposal for the entire ensemble of walkers at each step.

Even for a tool such as \emph{emcee}, it is impractical to employ the MCMC in the traditional fashion, where the full 34-dimensional parameter space would be sampled with density proportional to the likelihood.  Instead, we employ a novel iterative MCMC approach, dubbed ``fast burn-in'', designed for this analysis.  Rather than attempt to fill in the 34-dimensional volume, the ``fast burn-in'' aims to map the envelope of the likelihood function along multiple one-dimensional projections of the full parameter space.  The shape of each 1D envelope can then be used to constrain the projected parameter.

\begin{figure}[t!]
  \centering
  \includegraphics[width=0.48\textwidth]{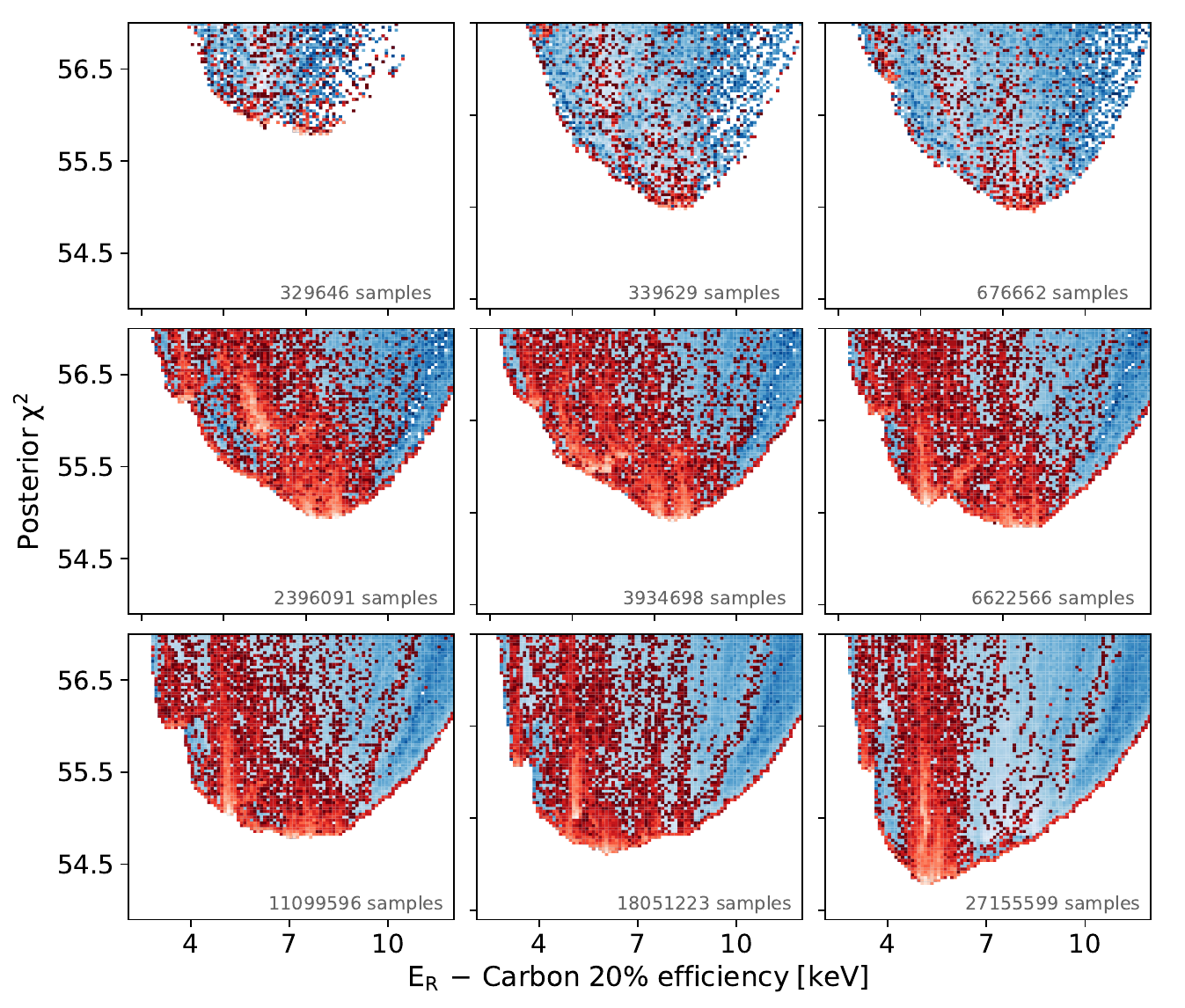}
    \caption{The sampled $\chi^2$ (or $-2\log\mathcal{L}$) distribution projected into $x_{\mathrm{C},20\%,2.45\mathrm{keV}}$ -- the carbon recoil energy with 20\% nucleation efficiency at a thermodynamic threshold of 2.45 keV -- showing the progressive sampling of the distribution over many epochs (blue) with the events selected to start the next epoch shown in red. The cumulative number of points evaluated is indicated in each frame.}
  \label{fig:convergence}
\end{figure}

To accomplish the fast burn-in, the MCMC is run in a series of fixed-length ``epochs''.  At the end of each epoch, the global set of likelihood evaluations thus far is examined, and points that fall on the boundary of the likelihood function in at least one of the 1D projections (i.e.\ points that yield the maximum likelihood observed so far at a given value of the parameter of interest) are selected as the starting positions for the next epoch's walker ensemble. In this case, the 1D projections are made on each of the 34 parameters of the likelihood function:  the $\{x_{s,p,T}\}$ and $\{s_k\}$.  That is, at the end of each epoch, all evaluations made thus far are projected onto the $(x,\log\mathcal{L})$ plane for each $x\in(\{x_{s,p,T}\}\cup\{s_k\})$.  Each projection is divided into $M$ bins in the projected dimension, and in each bin the point giving the highest $\log\mathcal{L}$ is identified.  This yields up to $34\times M$ points (selected points in each projection are not necessarily unique), which are used as the initial walker position set for the next epoch.  This process is illustrated in Fig.~\ref{fig:convergence}.

\begin{figure}[t!]
  \centering
  \includegraphics[width=0.48\textwidth]{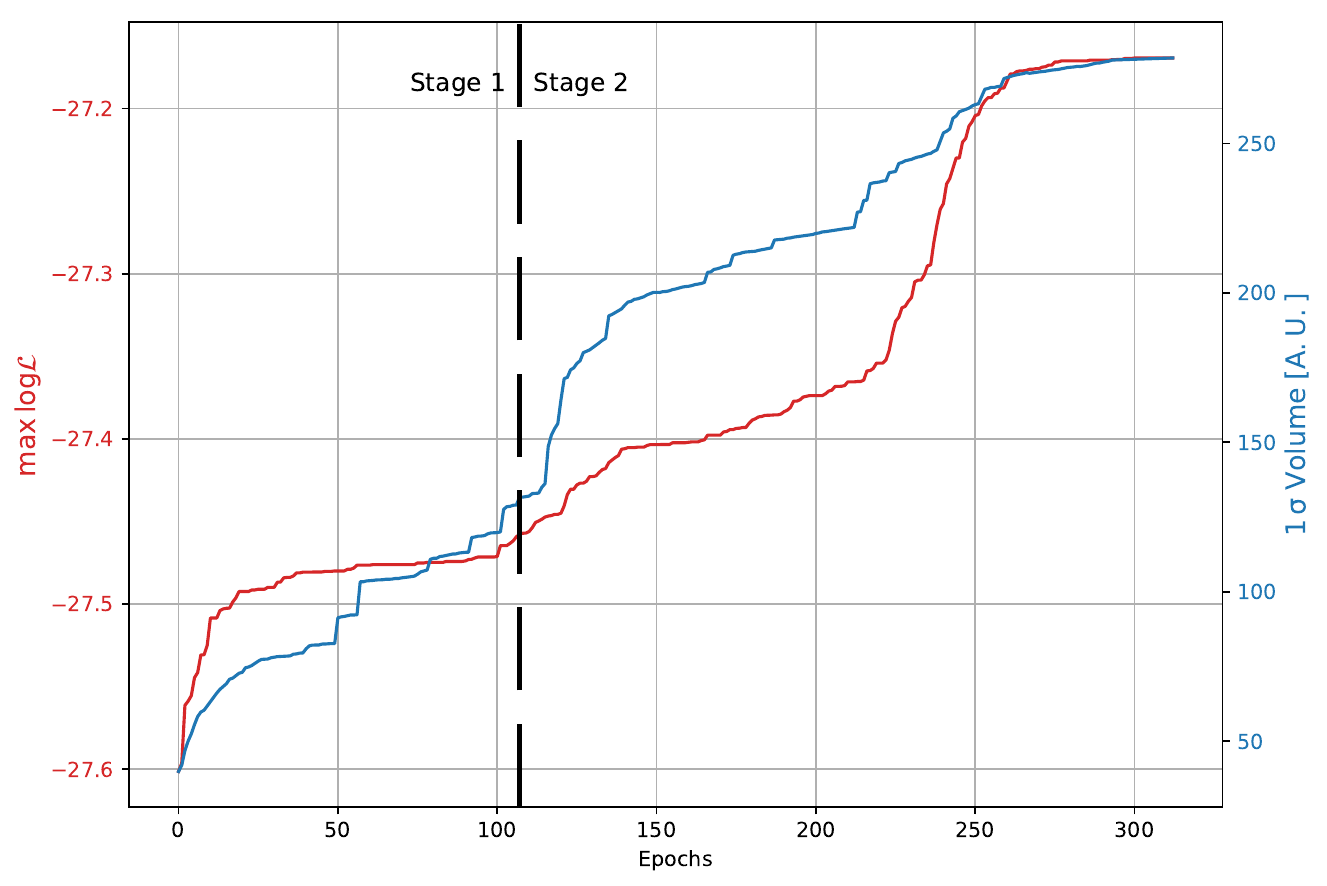}
  \caption{Progression of the maximum log-likelihood and ``$1 \sigma$'' volume of the likelihood function.}
  \label{fig:prog}
\end{figure}

This process can be further tuned by changing the proposal step scale of the MCMC (a unitless parameter defined in \emph{emcee}), the number of MCMC steps per epoch ($k$), and the number of bins for each parameter at the end of the epoch ($M$). The likelihood function was first mapped with a rapidly traversing ``exploratory'' phase with a proposal step scale of 2, $k$ = 5, and $M$ = 100. Then, a more thorough exploration was performed with a step scale of 1.2, $k$ = 10, $M$ = 500.

\begin{figure}[h!]
  \centering
  \includegraphics[width=0.48\textwidth]{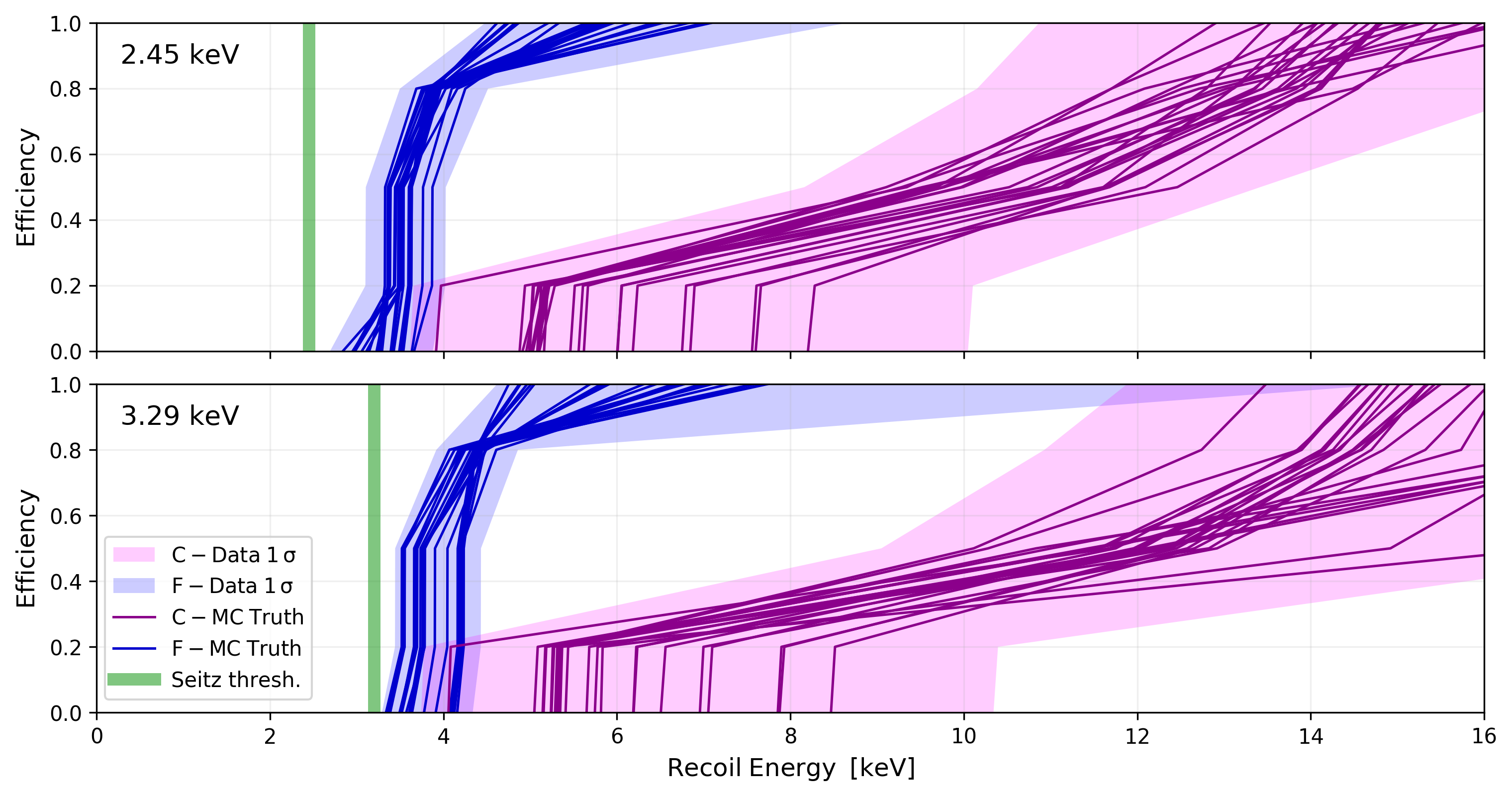}
  \caption{\label{fig:vmc_truth} Monte Carlo inputs for $25$ ``extended'' parametrically simulated datasets, compared to the $1 \sigma$ error bands from the fit to the real data.}
\end{figure}

For both phases of the MCMC exploration, the convergence criteria used were based on both the maximum likelihood reached and the ``volume'' of the likelihood function within  ``$1 \sigma$'' of the current best fit after each epoch. Specifically, the volume considered is the $34$-dimensional volume subtended by all MCMC samples within $0.5$ of the current maximum log-likelihood value, serving as a measure of the stability of the boundary of the likelihood function. Both of these criteria proved useful, as often the MCMC's progress would halt temporarily for one quantity but not the other. The progression is shown in Fig.~\ref{fig:prog}. The convergence criteria, checked after each epoch, was that there be $25$ consecutive epochs with less than a $0.1 \%$ change in log-likelihood or $1 \sigma$ volume. This criteria was validated using the Monte Carlo data described in Sec.\ \ref{S:MCstudy}. The MCMC fits of these simulated datasets were run until the aforementioned criteria was satisfied, plus an additional $50$ epochs in stage $2$ of the fit to catch any MCMCs potentially caught in local minima. No jumps in the maximum log-likelihood or $1 \sigma$ volume as previously defined were observed in the last $50$ epochs, suggesting that these criteria were sufficient.

\section*{\label{A:model bias}Appendix B: Model bias}

The fit results of 25 parametrically simulated datasets shown in Fig.~\ref{fig:bias} are all in reasonable agreement with the input parameters used to generate the Monte Carlo data (i.e.\ at no point do the fits completely lie outside the $1 \sigma$ band of the true model). However, it is also apparent from Fig.~\ref{fig:bias} that in some cases, there are persistent (albeit small) offsets between the average fit results and the input parameters; notably, the fluorine $100\%$ point is consistently overestimated, and the carbon efficiency curves underestimated. The mean residual for each parameter can be used to define a ``bias function'' $B_\theta$, with which the estimate of a parameter $\theta$ is given by:

\begin{equation}
    E[\hat{\theta}] = \theta_{\mathrm{True}} + B_\theta(\theta_{\mathrm{True}}).
    \label{eq:bias}
\end{equation}

\begin{figure}[t!]
  \centering
  \includegraphics[width=0.47\textwidth]{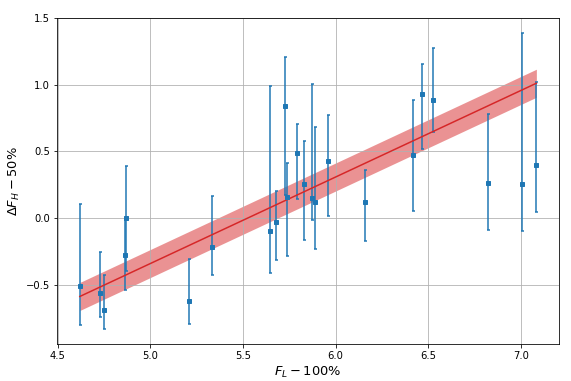}
  \caption{\label{fig:bias_func} Bias of the $50 \%$ efficiency point of fluorine at 3.29 keV vs.\ the true value of the fluorine $100\%$ efficiency point at 2.45 keV thermodynamic threshold. The Monte Carlo trial results are shown in blue, and a linear fit to the points with $1 \sigma$ error band are shown in red.}
\end{figure}

This can be used to amend reported uncertainties to account for biases inherent in the likelihood function, or even correct for them. Conservatively, one could expand the reported error bands for the data fit by subtracting the upper or lower (as appropriate) $1 \sigma$ limit of $B_\theta$ from each parameter. Less conservatively, one could shift the uncertainty band on both sides by the appropriate bias limit, and shift the reported best-fit value by the best-fit value of $B_\theta$.

However, this nominal method of application assumes that the bias function is ``flat'', i.e.\ that it is a constant not dependent on the true value of any parameters. Specifically, the presumption is that $B (\theta_\mathrm{True}) = B(\hat{\theta})$ for the real data. Since the Monte Carlo data described was generated from a single point in the parameter space, this assumption remains untested, with the worst-case scenario being that the bias function varies rapidly as a function of multiple parameters (the bias in $\theta_i$ could depend on the true value of $\theta_j$).

\begin{figure*}
  \centering
  \includegraphics[width=0.9\textwidth]{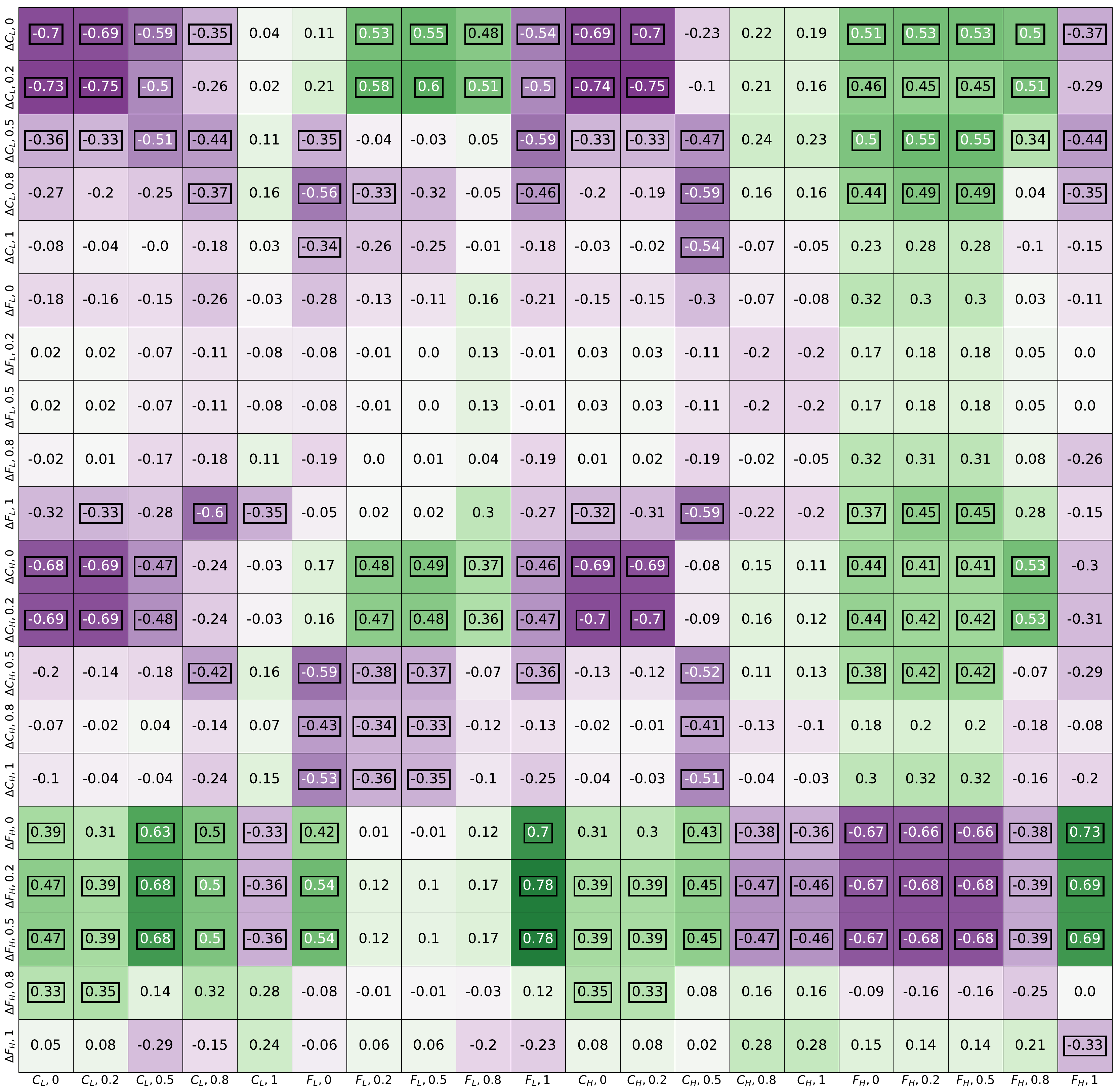}
  \caption{\label{fig:spearman} Spearman correlation coefficient for the bias of every parameter of interest (y axis) vs.\ the simulation input value (x axis). The color scale is indicative of this coefficient as well for easier visualization, and instances of a statistically significant correlation are boxed. Note that in this variable abbreviation, ``C'' or ``F'' indicate the target atom, ``L'' or ``H'' indicate the thermodynamic threshold fencepost as 2.45 or 3.29 keV respectively (``Low'' or ``High''), and the numeric value is the efficiency fencepost.}
\end{figure*}

To rectify this, ideally, datasets would be generated with each parameter varied independently to be $\pm\, 1 \sigma$ from the best-fit values, spanning the entire parameter space around the estimated best fit, with multiple trials in each case. However, this is not computationally practical. A more practical strategy is to instead vary all parameters around the best-fit point at once, by simulating datasets from efficiency curves drawn randomly from within the one sigma contour of the original fit to the data. $25$ such ``extended'' parametric Monte Carlo datasets were generated from the input efficiency curves shown in Fig.~\ref{fig:vmc_truth}. These datasets are fit in the usual way, showing biases with a similar magnitude and direction.

However, using the average residual to calculate the bias in each parameter as a function of itself alone ignores the possibility of dependence on other parameter true values. It is thus essential to examine the bias of each parameter of interest as a function of the true value of every parameter of interest (yielding $400$ combinations). Indeed, while there are many cases where the bias function does not vary with the input parameter, there are instances where the bias function is not constant. One such example is shown in Fig. \ref{fig:bias_func}.

One test that can succinctly summarize this complicated relationship is the correlation coefficient of the bias vs. true parameter value for every combination of the $20$ parameters of interest. Specifically, the Spearman correlation coefficient $\rho$ is appropriate in this case, as it is a nonparametric test \cite{spearman}. For a sample size of $25$, the null hypothesis that no correlations exist can be rejected at the $90\%$ confidence level if the $|\rho| > 0.324$ \cite{spearman}. This test statistic for every combination of the parameters of interest is shown in Fig.~\ref{fig:spearman}.

These results are then used to compute the bias function of every parameter, a 20-dimensional scalar function. A first or zeroth-order polynomial is used as a model for each parameter combination with or without a statistically significant correlation respectively. Before this, however, an outlier rejection procedure is applied to remove strenuous data points. For every parameter combination and every data point, a fit with a first-order polynomial is performed on all the data except the point in question. If this point is a $>\,2 \sigma$ outlier of that fit, it is rejected. For the final fit, only uncertainty in the y intercept is included to prevent the uncertainty of each bias function from vanishing at the central region of parameter space (due to uncertainty in the slope of the first-order fits). An example of one such linear fit and resulting error band is shown in Fig.~\ref{fig:bias_func}. This yields $400$ constraints on the true values of the $20$ parameters of interest in the form

\begin{equation}
    \hat{\theta}_j = \theta_j + B_{ji}(\theta_i)
\end{equation}

\begin{figure*}
  \center
  \includegraphics[width=0.4\textwidth]{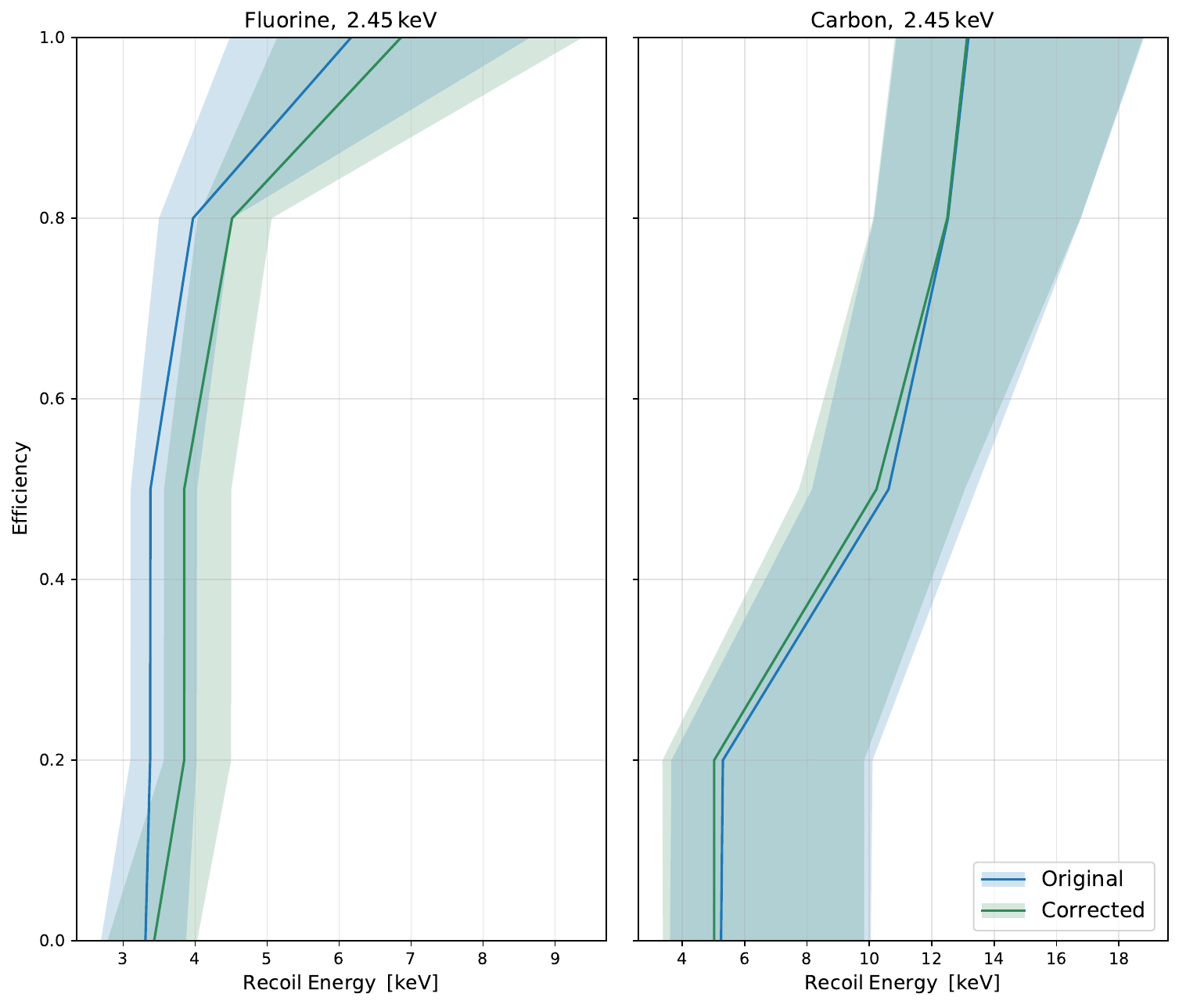}
  \includegraphics[width=0.4\textwidth]{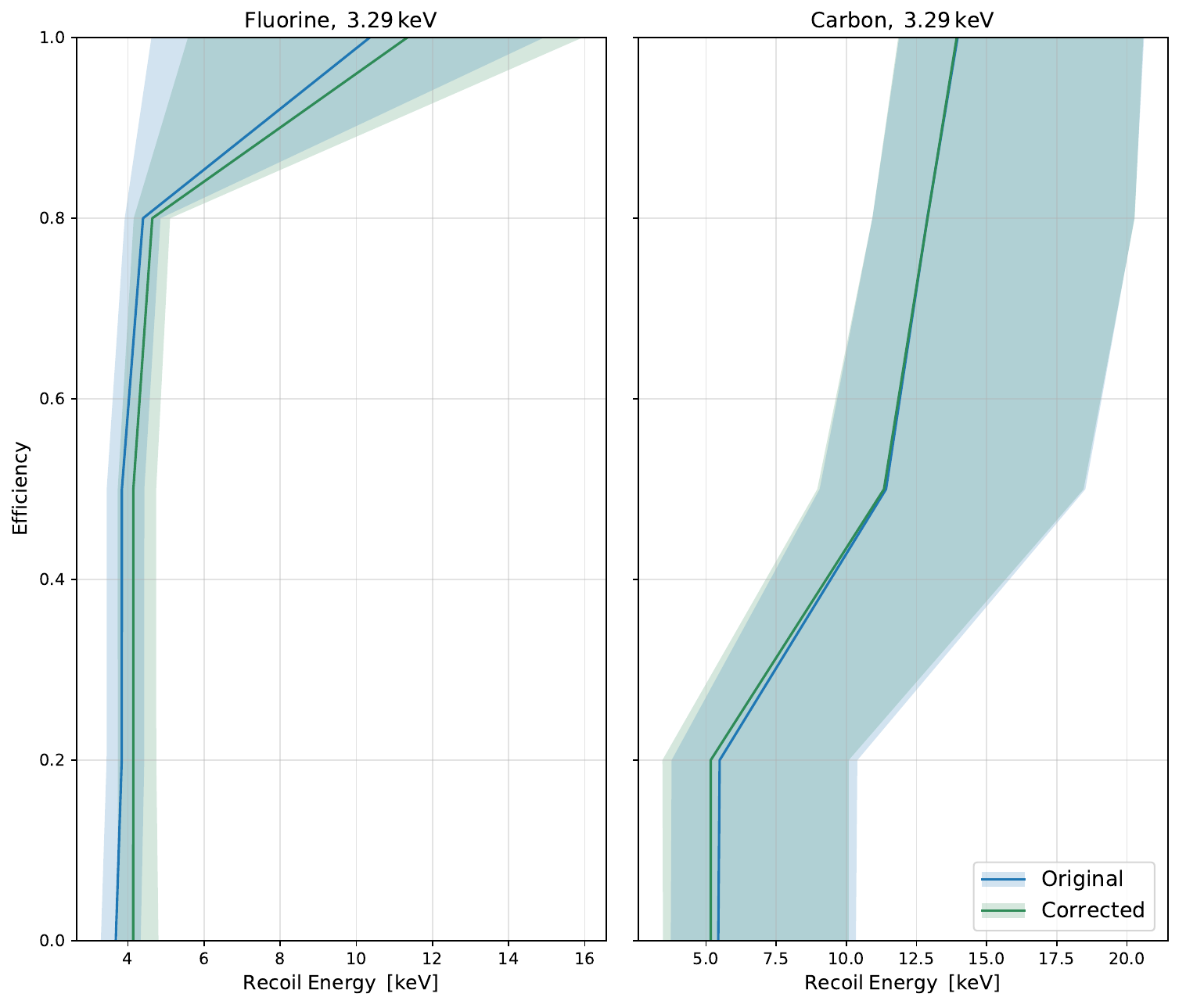}
  \caption{\label{fig:NFB} Original (blue) and bias-corrected (green) nucleation efficiency curves and $1 \sigma$ error bands for fluorine and carbon at thermodynamic thresholds of 2.45 keV (left) and 3.29 keV (right).}
\end{figure*}

\noindent relating one estimated parameter value $\hat{\theta}_j$ to the true values of the same parameter and another $\theta_i$, with one of the $400$ bias functions. $B_{ji}$ is the bias in parameter $j$ as a function of parameter $i$, and is normally distributed according to the first order polynomial fits described above. This is an overdetermined, nonlinear system of equations. To find an optimal solution, the following likelihood function was constructed and maximized:

\begin{equation}
    \log \mathcal{L} \left( \left \{\hat{\theta} \right \} | \left \{ \theta \right \} \right) = \sum_i \sum_j \log \left [ P_{\mathrm{Norm}} \left(\hat{\theta}_j = \theta_j + B_{ji} (\theta_i) \right) \right ].
\end{equation}

This fit was performed for the original best-fit parameter values to obtain the corrected best-fit nucleation efficiency curves. The new result is compared to the original efficiency curves in Fig.~\ref{fig:NFB}. It was not computationally practical to apply this method to all the MCMC samples explored in the original fit to build new $1 \sigma$ error bands. Therefore, the approximate new results presented in Fig.~\ref{fig:NFB} are produced by shifting the original error bands by the offset of the corrected best-fit curve and expanded by the statistical uncertainty in the fit of the corrected result. Fortunately, the bias-corrected result is not significantly discrepant with the original best-fit (within $1 \sigma$ agreement), so this new analysis does not cast doubt on the WIMP sensitivity results shown in previously published works using the same PICO calibration data (which are not bias corrected) \cite{pico60_v2, pico60_v3}. Indeed, the close overlap between the original and bias-corrected results justifies ignoring the fit biases in analyses with this data.

\bibliography{bibliography}

\end{document}